\documentclass[12pt,a4paper]{article}
\usepackage[dvips]{graphicx}
\usepackage{amssymb}
\usepackage{feynmp}

\makeatletter
\newif\if@preliminary
\@preliminaryfalse
\def\preliminary{\@preliminarytrue}
\def\preprintno#1{\def\@preprintno{#1}}
\def\address#1{\def\@address{#1}}
\def\email#1{\thanks{\tt #1}}
\def\abstract#1{\def\@abstract{#1}}
\renewcommand\abstractname{ABSTRACT}
\newlength\preprintnoskip
\newlength\abstractwidth
\@titlepagetrue
\renewcommand\maketitle{\begin{titlepage}%
  \let\footnotesize\small
  \def\thefootnote{\fnsymbol{footnote}}
  \hfill\parbox{\preprintnoskip}{%
  \begin{flushright}\@preprintno\end{flushright}}\hspace*{1cm}
  \vskip 60\p@
  \begin{center}%
    {\Large\bf\boldmath \@title \par}\vskip 1cm%
    {\sc\@author \par}\vskip 3mm%
    {\@address \par}%
    \if@preliminary
      \vskip 2cm {\large\sf (PRELIMINARY DRAFT) \par \@date}%
    \fi
  \end{center}\par
  \@thanks
  \vfill
  \begin{center}%
    \parbox{\abstractwidth}{\centerline{\abstractname}%
    \vskip 3mm%
    \@abstract}
  \end{center}
  \end{titlepage}%
  \setcounter{footnote}{0}%
  \def\thefootnote{\arabic{footnote}}
  \let\thanks\relax\let\maketitle\relax
  \gdef\@thanks{}\gdef\@author{}\gdef\@address{}%
  \gdef\@title{}\gdef\@abstract{}\gdef\@preprintno{}
}%
\def\@citex[#1]#2{\if@filesw\immediate\write\@auxout{\string\citation{#2}}\fi
  \def\@citea{}\@cite{\@for\@citeb:=#2\do
    {\@citea\def\@citea{,\penalty\@m}\@ifundefined
       {b@\@citeb}{{\bf ?}\@warning
       {Citation `\@citeb' on page \thepage \space undefined}}%
\hbox{\csname b@\@citeb\endcsname}}}{#1}}
\def\citerange{\@ifnextchar [{\@tempswatrue\@citexr}{\@tempswafalse\@citexr[]}}
\def\@citexr[#1]#2{\if@filesw\immediate\write\@auxout{\string\citation{#2}}\fi
  \def\@citea{}\@cite{\@for\@citeb:=#2\do
    {\@citea\def\@citea{--\penalty\@m}\@ifundefined
       {b@\@citeb}{{\bf ?}\@warning
       {Citation `\@citeb' on page \thepage \space undefined}}%
\hbox{\csname b@\@citeb\endcsname}}}{#1}}
\long\def\@makecaption#1#2{%
  \vskip\abovecaptionskip
  \sbox\@tempboxa{#1: \emph{#2}}%
  \ifdim \wd\@tempboxa >\hsize
    #1: \emph{#2}\par
  \else
    \hbox to\hsize{\hfil\box\@tempboxa\hfil}%
  \fi
  \vskip\belowcaptionskip}
\def\fmslash{\@ifnextchar[{\fmsl@sh}{\fmsl@sh[0mu]}}
\def\fmsl@sh[#1]#2{%
  \mathchoice
    {\@fmsl@sh\displaystyle{#1}{#2}}%
    {\@fmsl@sh\textstyle{#1}{#2}}%
    {\@fmsl@sh\scriptstyle{#1}{#2}}%
    {\@fmsl@sh\scriptscriptstyle{#1}{#2}}}
\def\@fmsl@sh#1#2#3{\m@th\ooalign{$\hfil#1\mkern#2/\hfil$\crcr$#1#3$}}
\makeatother

\def\fn{\footnotesize}

\newcounter{actr}

\newcommand{\fnnum}[1]{$^#1$}
\newcommand{\fncnum}[1]{$^{,#1}$}

\def\email#1{\relax}

\begin{document}
\begin{fmffile}{wwgraphs}
\preprintno{DESY--96--256 \\hep-ph/9708310\\ August 1997}
\title{%
 STRONGLY INTERACTING VECTOR BOSONS\\
 AT TeV $e^\pm e^-$ LINEAR COLLIDERS
}
\vspace{0.3cm}
\author{%
 E.~Boos\fnnum{1}\fncnum2\email{~Boos@ifh.de},
 H.--J.~He\fnnum3\email{~HJHe@mail.desy.de},
 W.~Kilian\fnnum4\email{~Kilian@thphys.uni-heidelberg.de},
 A.~Pukhov\fnnum2\email{~Pukhov@theory.npi.msu.su},
 C.--P.~Yuan\fnnum5\email{~Yuan@pa.msu.edu},\\[0.2cm]
 and P.M.~Zerwas\fnnum3\email{~Zerwas@desy.de}
}
\vspace{0.3cm}
\address{
 \fnnum1{}Technische Hochschule Darmstadt, Schlo\ss{}gartenstr.~9\\ 
 D--64289 Darmstadt, Germany\\[\baselineskip]
 \fnnum2{}Institute of Nuclear Physics, Moscow State University\\
 119899 Moscow, Russia\\[\baselineskip]
 \fnnum3{}Deutsches Elektronen-Synchrotron DESY\\
 D--22603 Hamburg, Germany\\[\baselineskip]
 \fnnum4{}Institut f\"ur Theoretische Physik, Universit\"at Heidelberg,
 Philosophenweg 16\\
 D--69120 Heidelberg, Germany \\[\baselineskip]
 \fnnum5{}Department of Physics and Astronomy, Michigan State
University\\
 East Lansing, Michigan 48824, USA
}
\abstract{%
In the absence of light Higgs bosons, the $W$ and $Z$ bosons become
strongly interacting particles at energies of about 1~TeV.  If the
longitudinal $W$,$Z$ components are generated by Goldstone modes
associated with spontaneous symmetry breaking in a new strong
interaction theory, the quasi-elastic $W$,$Z$ scattering amplitudes
can be predicted as a systematic chiral expansion in the energy.  We
study the potential of TeV $e^+e^-$ and $e^-e^-$ linear colliders in
investigating these scattering processes.  We estimate the accuracy
with which the coefficients of the chiral expansion can be measured in
a multi-parameter analysis.  The measurements will provide us with a
quantitative test of the dynamics underlying the $W$,$Z$ interactions.
}
\maketitle

\baselineskip20pt   
\section{Introduction}
\label{sec:intro}
Elastic scattering amplitudes of massive vector bosons grow
indefinitely with energy if they are calculated as a perturbative
expansion in the coupling of a non-abelian gauge theory.  As a result,
they manifestly violate unitarity beyond a critical energy scale
$\sqrt{s_c}$~\cite{Uni}.  In fact, the $S$-wave scattering amplitude
of longitudinally polarized $W,Z$ bosons in the isoscalar channel
$(2W^+W^-+ZZ)/\sqrt{3}$,
\begin{equation}
  a_0^0(s)  = \frac{\sqrt2\,G_F s}{16\pi} + O(g^2,g'^2)
\end{equation}
must be bounded by 1/2.  Unitarity therefore is violated for
energies in excess of
\begin{equation}
  \sqrt{s_c} \sim 1.2\;{\rm TeV}
\end{equation}
in $WW$ scattering.

This problem can be solved in two different ways.  In the Standard
Model~\cite{SM} a novel scalar particle, the Higgs boson, is
introduced to restore unitarity at high energies~\cite{Higgs,A1}.  The
additional contribution due to the exchange of this particle in the
scattering amplitude of longitudinal vector bosons cancels the
asymptotic rise of the Yang-Mills amplitude if the coupling of the
Higgs particle to the $W,Z$ bosons is chosen properly.  In that case,
the tree-level amplitude approaches a constant value.  Electroweak
observables in the fermion/gauge boson sector of the Standard Model
are affected by radiative corrections which depend logarithmically on
the Higgs boson mass $M_H$.  From the high-precision data at LEP1,
SLC, and the Tevatron, an upper limit of $M_H<550\;{\rm GeV}$ has been
derived at the $2\sigma$ level~\cite{MH-limits}.  This limit is not
sharp: Excluding one or two observables from the analysis weakens the
bound significantly~\cite{Baur}.  In a cautious conclusion the
experimental limit may therefore be interpreted within the minimal
model as indicative for a scale $<O(1\;{\rm TeV})$.

However, there exists a second solution to the unitarity problem.  If
the Higgs boson is not realized in Nature, the $W$ bosons become
strongly interacting particles at TeV energies.  In such a scenario
the experimental upper bound of $\sim 1\;{\rm TeV}$ can be
re-interpreted as the cut-off scale up to which the Standard Model of
fermions and vector bosons may be extended before new physical
phenomena become apparent.  Such novel strong interactions of the $W$
bosons may be indicated by slight deviations of the static electroweak
$W,Z$ parameters from the predictions in the Standard Model,
\emph{i.e.}, for the oblique parameters, the $Z$-fermion couplings,
the magnetic dipole, and the electric quadrupole moments of the
$W^\pm$ bosons~\cite{STU,TGV,CCD}.  However, besides the production of
triple gauge bosons in $e^+e^-$ annihilation~\cite{3W}, the classical
test ground for these interactions is the elastic and quasi-elastic
$2\to 2$ scattering experiments of the $W^\pm$ and $Z$ bosons
\begin{equation}
  WW\to WW
\end{equation}
where $W$ generically denotes the particles $W^\pm, Z$.

It is natural, though not compulsory, to trace back the strong
interactions of the $W$ bosons to a new fundamental strong interaction
characterized by a scale of order $1\;{\rm TeV}$~\cite{Mass-gen}.  If
the Lagrangean of the underlying theory is globally chiral-invariant,
this symmetry may be broken spontaneously.  The Goldstone bosons
associated with the spontaneous symmetry breaking can be absorbed by
the gauge bosons to generate the masses and to build up their
longitudinal degrees of freedom.  It may be assumed in this scenario
that the breaking pattern of the chiral symmetry in the strongly
interacting sector is such that $SU(2)\times SU(2)\to SU(2)_c$ leaves
the isospin group $SU(2)_c$ unbroken.  This custodial $SU(2)_c$
symmetry~\cite{Mass-gen} automatically ensures that the
$\rho$~parameter, the ratio of the neutral-current to charged-current
couplings, is unity up to small perturbative corrections. This
condition~\cite{SU2c} is strongly supported by the electroweak
precision data.  The fact that in such a scenario the longitudinally
polarized $W$ bosons are associated with the Goldstone modes of chiral
symmetry breaking, has far-reaching consequences which are formalized
in the Equivalence Theorem~\cite{A1,ET,He,ETadd}.  This mechanism can be
exploited to predict the scattering amplitudes of the $W_L$ bosons for
high energies below the mass scale of new resonances\footnote{This is
the analog to low-energy pion physics below the $\rho$~resonance of
QCD, in which the pions are the Goldstone bosons associated with the
spontaneous chiral $SU(2)\times SU(2)$ symmetry breaking.}.  Expanding
the scattering amplitudes in powers of the energy $\sqrt{s}$, the
leading term is parameter-free, thus being a consequence \emph{per se}
of the chiral symmetry breaking mechanism, independent of the
particular dynamical theory.  The higher-order terms in the chiral
expansion depend on new coefficients which reflect the detailed
structure of the underlying strong-interaction theory.  With rising
energy they may evolve towards a resonant behavior, in the scalar or
vector channels for instance.

To study potentially strong interactions between $W$ bosons requires
energies in the TeV range.  They will be provided by the $pp$ collider
LHC and by future $e^+e^-$ linear colliders which will operate in the
second phase at energies of $1.5$ to $2\;{\rm TeV}$, see \emph{e.g.}
Ref.\cite{LC}.  Longitudinal $W$ bosons are radiated off quarks and
electrons/positrons with a probability $g^2/16\pi^2\sim 3\times
10^{-3}$; since the $Z$ charge of leptons is small, the radiation of
$Z$ bosons is suppressed compared to $W$ bosons.  The following
(quasi-)elastic processes can be studied in $e^+e^-$ and
$e^-e^-$ collisions~\cite{BCHP,Cuypers,eminus}:
\begin{equation} \label{WW}
  \begin{array}{l@{\quad\to\quad}l@{\quad:\quad}l@{\quad\to\quad}l}
  e^+e^- & \bar\nu_e\nu_eW^+W^- & W^+W^- & W^+W^-\\
  e^+e^- & \bar\nu_e\nu_eZZ     & W^+W^- & ZZ\\
  e^-e^- & \nu_e\nu_eW^-W^-     & W^-W^- & W^-W^-
  \end{array}
\end{equation}
It turns out that the rates for these processes are sufficiently large
for thorough analyses at $e^+e^-$ c.m.\ energies of $\sqrt{s}\sim
1\;{\rm TeV}$ and above.  Other processes involving initial
state $Z$~bosons,
\begin{equation}\label{WZ}
  \begin{array}{l@{\quad\to\quad}l@{\quad:\quad}l@{\quad\to\quad}l}
  e^+e^- & \bar\nu_e e^-W^+Z & W^+Z & W^+Z\\
  e^+e^- & e^+e^-ZZ          & ZZ   & ZZ
  \end{array}
\end{equation}
are suppressed for the reasons discussed above.  Nevertheless, they
must be investigated to achieve a complete determination of the
quartic gauge interactions in next-to-leading order of the chiral
expansion.  Since all basic scattering processes
(\ref{WW}) and (\ref{WZ}) lead to different final states, they can be
disentangled in principle [though this may not be so straightforward
in practice since the final state electrons and positrons may be lost
in the forward directions].

The main objective of the present analysis are theoretical predictions
for the processes (\ref{WW}) and (\ref{WZ}) in the region where the
$W,Z$ bosons become strongly interacting but the energies do not reach
yet the resonance region, which may be delayed until a scale of $4\pi
v\sim 3\;{\rm TeV}$ is approached.  We study the predictions in
leading order of the chiral expansion and analyze the sensitivity to
next-to-leading order contributions\footnote{Preliminary results of
this study have been presented in Ref.\cite{proc}.}. This will
enable us to estimate the accuracy with which the parameter-free
leading-order amplitudes can be measured.  If the Higgs mechanism is
not realized in Nature, these analyses will shed light on the symmetry
structure and the basic physical mechanism that provides masses to the
fundamental electroweak bosons.  Alternative approaches that are not
based on chiral symmetry breaking, would in general lead to quite
different predictions for $WW$ scattering amplitudes.

The paper is organized as follows. In Sec.\ref{sec:chiral} we briefly
recapitulate the basic formalism of electroweak chiral Lagrangeans.
In Sec.\ref{sec:WW} the helicity amplitudes for the $WW\to WW$ fusion
signals are analyzed, while Sec.\ref{sec:IEWA} is devoted to the
equivalent particle approximations and kinematical improvements.  This
discussion serves as a useful guideline for the analysis and as an
independent check for the complete 
$f_1 f_2 \rightarrow {f'}_1{f'}_2 WW$
 tree-level calculations.
The full calculation and the results for probing both the custodial
$SU(2)_c$ conserving and breaking chiral parameters at TeV $e^\pm e^-$
linear colliders are presented in Sec.\ref{sec:comphep}
and~\ref{sec:SU2c}.  Conclusions are given in Sec.\ref{sec:conc}.  In
Appendices~\ref{sec:uni} and~\ref{sec:loop}, constraints from
unitarity bounds are derived and the leading contributions of the
one-loop radiative corrections are estimated.  The exact tree-level
$WW\to WW$ helicity amplitudes are summarized in compact form up to
next-to-leading order in Appendix~\ref{sec:hel}.

\section{Chiral Lagrangeans}\label{sec:L}
\label{sec:chiral}
For theories in which the chiral symmetry is broken spontaneously,
\emph{i.e.}, $SU(2)\times SU(2)\to SU(2)_c$, effective Lagrangeans can
be defined for the associated Goldstone fields.  They correspond to
expansions in the dimensions of the field operators, or equivalently
in the energy $\sqrt{s}$ in momentum space~\cite{ChPT,EWChPT}.  This
systematic expansion leads to a parameter-free leading-order
interaction in the Lagrangean, supplemented by higher-order terms
which reflect the detailed structure of the underlying strong
interaction theory.  Thus the leading-order interaction is a direct
model-independent consequence of chiral symmetry breaking \emph{sui
generis}.  The Equivalence Theorem then allows to re-interpret
scattering amplitudes derived for the Goldstone particles as
equivalent to the scattering amplitudes of the longitudinally
polarized $W,Z$ particles for asymptotic energies $E(W,Z)\gg M_{W,Z}$.

The kinetic terms of the gauge fields and the first terms in the
chiral Lagrangean of the Goldstone fields are given by the following
expansion:
\begin{eqnarray}\label{chL}
  {\cal L} &=& {\cal L}_g + {\cal L}_e \nonumber\\
  && {}+ {\cal L}_0 + {\cal L}_4 + {\cal L}_5 + \ldots
\end{eqnarray}
${\cal L}_g$ denotes the kinetic terms of the $W^{\pm,3}$ and $B$ 
fields\footnote{The complete Lagrangean is understood to contain
the usual gauge-fixing and ghost terms.}.  The $SU(2)\times U(1)$
gauge fields are coupled to the matter fields through covariant
derivatives in ${\cal L}_e$.  These two parts of the Lagrangean are
given by the expressions
\begin{eqnarray}
  {\cal L}_g &=& -\textstyle\frac18{\rm tr}[W_{\mu\nu}^2]
                - \textstyle\frac14 B_{\mu\nu}^2\\
  {\cal L}_e &=& 
                \bar e_{\rm L}iD\!\!\!\!/\,\,e_{\rm L} 
                + ({\rm L}\leftrightarrow{\rm R})
\end{eqnarray}
with the usual definition of the covariant $SU(2)\times U(1)$
derivative in terms of the vector fields, the $SU(2)$ generators
$T^a$, and the hypercharge $Y$:
\begin{equation}
  iD_\mu = i\partial_\mu + g\vec T \cdot\vec W_\mu - g' {Y \over 2} B_\mu
\end{equation}
where $2 {\vec T}$ is equal to the Pauli matrix $\vec \tau$.
In the general $R_\xi$ gauge the Goldstone fields are described by the
unitary matrix\footnote{In the Standard Model, $U$ is the Goldstone
boson matrix which generates the Higgs isodoublet field from the real
Higgs field in the $R_\xi$ gauges.}
\begin{equation}
  U = \exp[-i{\vec w}\cdot{\vec\tau}/F]
\end{equation}
The custodial-symmetric dimension-2 operator of the Goldstone fields
is then given by
\begin{equation}
  {\cal L}_0 =  \textstyle\frac{F^2}{4}{\rm tr}[D_\mu U^\dagger D^\mu U]
\end{equation}
The coupling between the Goldstone particles and the $W$, $B$ gauge
fields is parameterized by the coefficient $F$.  The value of this
parameter is fixed by the measured $W$ or $Z$ masses,
\begin{equation}
  {\cal L}_0 =  M_W^2 W^+ W^- + \frac12 M_Z^2 Z^2 +\ldots
\end{equation}
so that the experimental value
\begin{equation}
  F = (\sqrt2\,G_F)^{-1/2} = 246\;{\rm GeV}
\end{equation}
can be derived for $F$ from the Fermi constant.  In the Standard
Model, $F$ is replaced by the expectation value $v$ of the Higgs field
in the ground state, $F=v$.  However, the physical interpretation of
these parameters is completely different in the two
scenarios\footnote{From now on, we will nevertheless adopt the
symbol~$v$ to characterize the weak-interaction scale, as generally
done in the literature.}.

A vector field $V_\mu$ can be defined by the Goldstone fields as
\begin{equation}
  V_\mu = U^\dagger D_\mu U
\end{equation}
corresponding to the derivative $\partial_\mu \vec w+\ldots$ for small
field strengths.  From the vector field two independent dimension-4
operators may be formed
\begin{eqnarray}
\label{L4}
    {\cal L}_4 &=& \alpha_4\,
                        {\rm tr}\left[V_\mu V_\nu\right]\,
                        {\rm tr}\left[V^\mu V^\nu\right] \\
\label{L5}
    {\cal L}_5 &=& \alpha_5\,
                        {\rm tr}\left[V_\mu V^\mu\right]\,
                        {\rm tr}\left[V_\nu V^\nu\right]
\end{eqnarray}
which describe the first two non-leading and model-dependent terms in
the chiral expansion.  The two interaction terms ${\cal L}_4$ and
${\cal L}_5$ are custodial symmetric, leaving the value $\rho=1$
unchanged.  Since they involve at least a quartic coupling of the
Goldstone particles, they affect in lowest order only $2\to2$
scattering processes but do not affect the trilinear vertices.  Thus,
$\alpha_4$ and $\alpha_5$ can only be determined in $WW\to WW$
scattering.  [Additional dimension-4 operators affect the trilinear
couplings; in this analysis they are assumed to be pre-determined by
standard methods such as $WW$ pair production in $e^+e^-$
annihilation.]

We assume that all higher-order coefficients in the chiral expansion
are much smaller than unity.  Even though a gauge-symmetric chiral
Lagrangean can be defined formally for any theory with a particular
particle content, this is meaningful only if the chiral series can be
truncated at a fixed operator dimension ($d=4$ for our purpose) and
still higher orders can be neglected.  However, if the concept of
spontaneous chiral symmetry breaking were not realized in Nature,
higher-order coefficients would be so large that an infinite number of
terms would enter even at the $W,Z$ mass scale.  In that case, the
above effective-theory formalism must be abandoned.

From the magnitude of loop effects which carry a factor $1/16\pi^2$
together with an additional power of $s/v^2$, the largest value of
$\sqrt{s}$ for a chiral expansion to be valid may be
estimated~\cite{NDA} as $\sqrt{s}\lesssim 4\pi v\sim 3\;{\rm TeV}$.
Thus, if the coefficients $\alpha_i$ in the chiral expansion were
experimentally required to be substantially larger than $1/16\pi^2$,
new resonance effects would already appear below the $3\;{\rm TeV}$
scale, \emph{e.g.}, thresholds for resonance production would become
visible in the intermediate range between about $1$ and $3\;{\rm
TeV}$.

Although the 't~Hooft-Feynman gauge turns out to be most convenient
for the computation method described below (Sec.\ref{sec:comphep}),
all observable quantities can be calculated equally well within the
unitary gauge in which the Goldstone fields $\vec w$ are set to zero.
In this gauge the physical content of the various terms becomes more
transparent: The standard vector boson interactions are determined by
the Yang-Mills kinetic Lagrangean alone, ${\cal L}_0$ just provides
the $W,Z$ masses, and the new dimension-4 operators ${\cal L}_{4,5}$
are recognized as two independent contact-interaction terms for the
$W,Z$ vector bosons:
\begin{eqnarray}
  {\cal L}_0 &=& 
        M_W^2 W^+_\mu {W^-}^\mu + \frac12 M_Z^2 Z_\mu Z^\mu\\
  {\cal L}_4 &=& \alpha_4\left[
        \frac{g^4}{2}\left[(W^+_\mu {W^-}^\mu)^2 
	+ (W^+_\mu{W^+}^\mu)(W^-_\nu{W^-}^\nu)\right]
	\right.\nonumber\\
	&& {}\left.
        + \frac{g^4}{c_w^2}(W^+_\mu Z^\mu)(W^-_\nu Z^\nu) 
	+ \frac{g^4}{4c_w^4}(Z_\mu Z^\mu)^2
        \right]\\
  {\cal L}_5 &=& \alpha_5\left[
        {g^4}(W^+_\mu{W^-}^\mu)^2 
	+ \frac{g^4}{c_w^2}(W^+_\mu{W^-}^\mu)(Z_\nu Z^\nu)
        + \frac{g^4}{4c_w^4}(Z_\mu Z^\mu)^2
        \right]
\end{eqnarray}
[$c_w^2=1-\sin^2\theta_w$ and $g^2=e^2/\sin^2\theta_w$].  The contact
terms introduce all possible quartic couplings $W^+W^-W^+W^-$,
$W^+W^-ZZ$, and $ZZZZ$ among the weak gauge bosons, that are
compatible with charge conservation and custodial $SU(2)_c$ symmetry.

\section{$WW$ scattering}
\label{sec:WW}
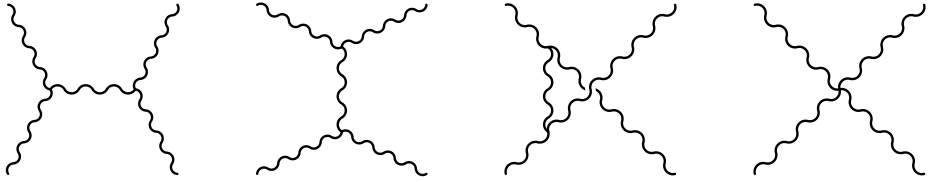
\begin{figure}
\begin{center}
\unitlength1mm
\begin{fmfgraph}(30,30)
  \fmfsurroundn{e}{8}
  \fmf{boson}{e4,v1,v2,e2}\fmf{boson}{e6,v1}\fmf{boson}{v2,e8}
\end{fmfgraph}
\begin{fmfgraph}(30,30)
  \fmfsurroundn{e}{8}
  \fmf{boson}{e4,v1,v2,e6}\fmf{boson}{e2,v1}\fmf{boson}{v2,e8}
\end{fmfgraph}
\begin{fmfgraph}(30,30)
  \fmfsurroundn{e}{8}
  \fmf{phantom}{e4,v1,e2}\fmf{phantom}{e6,v1,e8}\fmffreeze
  \fmf{phantom}{e4,v3,v1}\fmf{phantom}{e6,v4,v1}\fmffreeze
  \fmf{boson}{e4,e8}\fmf{boson,rubout=4}{v4,e2}
  \fmffreeze
  \fmf{boson}{e6,v4,v3}
\end{fmfgraph}
\begin{fmfgraph}(30,30)
  \fmfsurroundn{e}{8}
  \fmf{boson}{e4,v1,e2}\fmf{boson}{e6,v1,e8}
\end{fmfgraph}
\end{center}
\caption{Feynman graphs for (quasi-)elastic $WW$ scattering.}
\label{WW-graphs}
\end{figure}

From the effective chiral Lagrangean, the $2\to 2$ (quasi-)elastic
$WW$ scattering amplitudes can easily be derived.  As shown
generically in Fig.\ref{WW-graphs}, they involve $s$-channel,
$t/u$-channel exchange diagrams, and the non-abelian quartic boson
coupling, with their sum growing asymptotically proportional to $s$.
The additional quartic contributions introduced by ${\cal L}_4$ and
${\cal L}_5$ rise proportional to $s^2$.  The maximal power of $s$ is
realized only for amplitudes in which all four vector bosons are
longitudinally polarized; replacing any longitudinally polarized
external particle by a transversely polarized particle removes one
factor of $\sqrt{s}/v$; at the same time an additional power of the
weak couplings $g,g'$ is introduced.  [In the extreme forward and
backward directions where $t,u$ are of the order $M_{W,Z}^2$, the
power counting is invalid and both longitudinal and transversal
degrees of freedom contribute with comparable magnitude.]

It follows~\cite{Uni,ET} from analyticity, crossing symmetry, CP
invariance, and custodial symmetry, that to leading order in the
Yang-Mills couplings all (quasi-)elastic amplitudes can be expressed
in terms of a single function $A(s,t,u)$ which is symmetric with
respect to the exchange $(t\leftrightarrow u)$.  This function is
analytic in the Mandelstam variables $s,t,u$ apart from the usual
one-particle pole and two-particle cut singularities.  The Mandelstam
variables are given by the total energy and the momentum transfer in
the scattering processes: $s=E_{c.m.}^2$, $t(u)\approx
-s(1\mp\cos\theta)/2$ for $|s|,|t|,|u|\gg M_W^2$.  The amplitudes of
the scattering processes (\ref{WW}) and (\ref{WZ}) can be derived from
the master amplitude~$A$ in the following way:
\begin{eqnarray}
  A(W^+W^-\to ZZ) &=& A(s,t,u) \label{aZZ}\\
  A(W^+W^-\to W^+W^-) &=& A(s,t,u) + A(t,s,u) \label{aWW}\\
  A(W^-W^-\to W^-W^-) &=& A(t,s,u) + A(u,t,s) \label{asWW}
\end{eqnarray}
and
\begin{eqnarray}
  A(W^+Z\to W^+Z) &=& A(t,s,u) \label{aWZ}\\
  A(ZZ\to ZZ) &=& A(s,t,u) + A(t,s,u) + A(u,t,s) \label{a4Z}
\end{eqnarray}
To leading order in the energy expansion the amplitude $A(s,t,u)$ is
reduced to the simple expression
\begin{equation}
  A(s,t,u)_{\rm LO} = \frac{s}{v^2}
\end{equation}
which is parameter-free.  The next-to-leading order terms modify this
result, and the final tree-level expression is given to order $s^2$ by
\begin{equation}\label{aZZ-expr}
  A(s,t,u) = \frac{s}{v^2} + \alpha_4\frac{4(t^2+u^2)}{v^4}
        + \alpha_5\frac{8s^2}{v^4}
\end{equation}
The relations (\ref{aZZ}--\ref{a4Z}) for the amplitudes are preserved
by loop corrections and they are valid to all orders for
chirally-symmetric strong interactions.  There are, however,
additional perturbative corrections which are proportional to the
Yang-Mills couplings $g,g'$, with the $g'$ coupling breaking the
custodial symmetry.  Amplitudes involving transversely polarized
vector bosons, which are subleading both for high energies and in the
weak coupling expansion, do not respect the relations
(\ref{aZZ}--\ref{a4Z}).

It is instructive to analyze the angular momentum states that are
populated in $WW$ scattering.  The helicity analysis~\cite{Hel} of the
scattering amplitudes leads to the following decomposition in the
angular momentum
\begin{equation}
  A(00,00) = \sum_J A_J(00,00)\;d_{00}^J(\theta)
\end{equation}
for longitudinally polarized vector bosons, where
$d_{00}^J=P_J(\cos\theta)$ are the Legendre Polynomials.

Choosing the process $W^+W^-\to ZZ$ for example, the gauge
contributions to the amplitudes involve $t$- and $u$-channel exchange
diagrams, giving rise to arbitrarily high orbital angular momentum
states.  Therefore we decompose the amplitude with respect to spin
only, \emph{i.e.}, the residues of the poles for $t/u$-channel
diagrams are expanded:
\begin{equation}
  A_J = \displaystyle\frac{s^2}{4M_W^4}\left[{g^2c_w^4}\left(
        \frac{s}{2(t-M_W^2)}\hat A_t + \frac{s}{2(u-M_W^2)}\hat A_u
        + \hat A_c\right)
        + g^4\left(\alpha_4\hat A_4 + \alpha_5\hat A_5\right)\right]
\end{equation}
The subscripts $t,u,c$ for $\hat A$ denote the $t,u$ exchange and the
four-boson contact terms, respectively
(Tab.\ref{tab:hel-ZZ})\footnote{For the process $W^+W^-\to W^+W^-$,
the complete decomposition is given in the Appendix.}.

\begin{table}
\newcommand{\strt}{\rule[-5pt]{0pt}{18pt}}
\begin{displaymath}
\begin{array}{|r|rrr|cc|}
\hline
&\hat A_t & \hat A_u & \hat A_c & \hat A_4 & \hat A_5 \strt\\
\hline
J=0 & -\frac{20}{3} & -\frac{20}{3} & -\frac{16}{3} & \frac{8}{3} & 8
\strt\\
1 & \frac{44}{5} & -\frac{44}{5} & 0 & 0 & 0 \strt\\
2 & -\frac{4}{3} & -\frac{4}{3} & \frac{4}{3} & \frac{4}{3} & 0 \strt\\
3 & -\frac{4}{5} & \frac{4}{5} & 0 & 0 & 0 \strt\\
\hline
\end{array}
\end{displaymath}
\caption{Amplitude decomposition for the process $W^+W^-\to ZZ$ in the
limit $E\gg M_W$.}
\label{tab:hel-ZZ}
\end{table}

In the spin amplitudes, the contact term contains angular momenta
$J=0$ and $2$.  In the $t/u$ channel diagrams the additional vector boson
in the intermediate state populates,
together with the external vector bosons,
the states up to $J=3$.  In the limit $|s|,|t|,|u|\gg M_W^2$ the
leading $s^2$ behavior cancels for $\alpha_4=\alpha_5=0$; however, in
the forward/backward regions ($|t|,|u|\sim M_W^2$) this cancellation
needs not occur.  In other processes such as $W^+W^-\to W^+W^-$ there
is an additional $s$-channel diagram which is purely spin-1, since a
single vector boson $Z/\gamma$ is exchanged.

Given the helicity amplitudes, the differential cross sections can be
written as
\begin{equation}
  \frac{d\sigma}{d\cos\theta}
  (W_{\lambda_1}W_{\lambda_2}\to W_{\lambda_3}W_{\lambda_4}) 
  = 
  \frac{1}{32\pi s}
  \left| A(\lambda_1\lambda_2,\lambda_3\lambda_4)\right|^2~.
\end{equation}
This cross section can easily be integrated over all angles,
\begin{equation}
  \sigma(W_{\lambda_1}W_{\lambda_2}\to W_{\lambda_3}W_{\lambda_4}) 
  = 
  \frac{\eta}{32\pi s} 
  \int_{-1}^1 d(\cos\theta )
  \left| A(\lambda_1\lambda_2,\lambda_3\lambda_4)\right|^2
\end{equation}
where $\eta=\frac12(1)$ accounts for (non-)identical particles in the
final state.

Even though the longitudinal helicities build up the asymptotically
leading cross section $\sigma(W_LW_L\to W_LW_L)$, it cannot be
identified with the total cross section without applying angular cuts
for non-asymptotic energies since the forward peak for the scattering
of transversely polarized $W$ bosons gives rise to additional large
contributions to the total cross section.

Interference effects between different helicity amplitudes in the
initial state have to be taken into account in the non-asymptotic
regime.  Since the $W$ bosons are radiated off the electrons and
positrons, a coherent mixture of $W_{\lambda_1}W_{\lambda_2}$ helicity
states is generated with $\lambda_1$ and $\lambda_2=\pm,0$.
Interference effects in the final $W_{\lambda_3}W_{\lambda_4}$ state
need only to be included if the angular and energy distributions of the
leptons or jets in the  $W_3,W_4$ decays are analyzed explicitly.

\section{Equivalent particle approximations}
\label{sec:IEWA}
The elastic scattering of $W$~bosons at high energies will be studied
in TeV $e^+e^-$ and $e^-e^-$ collisions.  At high energies
electron/positron beams split for a long time into (neutrino $+W$) or
(electron/positron $+Z$) pairs.  In fact, if the transverse momentum
in the splitting process is $p_\perp$, the lifetime of the split state
is of order $\tau\sim E_e/(p_\perp^2+M_W^2)$ in the laboratory, which
is large for high electron/positron energies.  With $E_e=800\;{\rm
GeV}$ the lifetime $\tau\sim 10^{-1}\;{\rm GeV}^{-1}$ is an order of
magnitude longer than the weak interaction scale $\tau_w\sim
M_W^{-1}\sim 10^{-2}\;{\rm GeV}^{-1}$.  The $W$~bosons can therefore
be approximately treated as equivalent particles~\cite{EWA}, similar
to the equivalent photon approximation in QED~\cite{EPA}.  Moreover,
the splitting probability is maximal for small transverse momenta
$p_\perp\lesssim M_W$.  In the final picture, the $W$~bosons can be
treated as real particle beams which accompany the parent $e^\pm$
beams in the accelerator.

The energy spectrum of the $W$~bosons can conveniently be determined,
in the spirit of the discussion above, by old-fashioned perturbation
theory~\cite{OPT}.  Denoting the fraction of energy transferred from
the initial lepton to the $W$~boson by $x$, with $0\leq x\leq 1$, 
the spectra, under the leading logarithmic approximation, are given by~\cite{EWA}:
\begin{enumerate}
\item\emph{Transversely polarized $W^\pm$ bosons:}
\begin{eqnarray}\label{fT(x)-LLA}
  f^T_{W/e}(x) &=& \frac{\alpha}{4\pi s_w^2}\,
        \frac{1+(1-x)^2}{2x}\ln\frac{\hat s}{M_W^2}
\end{eqnarray}
where $\hat s=xs$.

For $e^-$~beams, the term $\sim 1$ corresponds to negative helicity of
the $W$~boson, while the term $\sim (1-x)^2$ corresponds to positive
helicity, suppressed for $x \to 1$ by the conservation of angular
momentum.  [The role of the helicities is interchanged for
$e^+$~beams.]  The spectrum increases with the logarithm of the
energy, which is a consequence of the unlimited transverse momentum of
the point-like coupling in the splitting process.
\item\emph{Longitudinally polarized $W^\pm$ bosons:}
\begin{eqnarray}
  f^L_{W/e}(x) &=& \frac{\alpha}{4\pi s_w^2}\,
        \frac{1-x}{x}
\end{eqnarray}
Since the emission of longitudinally polarized $W$~bosons is
suppressed for large transverse momentum, the longitudinal spectra are
not logarithmically enhanced.
\end{enumerate}

In the equivalent particle approximation the cross section $d\sigma$
for the colliding beam process, such as
\begin{equation}
  e^+e^-\to \bar\nu_e\nu_eW^+W^-
  \qquad\mbox{via}\qquad
  W^+W^-\to W^+W^- 
\end{equation}
can be obtained by convoluting the cross section $d\hat\sigma$ of the
$WW$ subprocess with the spectra of the two initial-state
$W$~bosons:\footnote{
A formalism, improved further, but with more complexity, can
be found in Ref.\cite{kuss}.} 
\begin{eqnarray}
  \lefteqn{d\sigma[e^+e^-\to \bar\nu_e\nu_eW^+W^-]}
  \nonumber\\
  &=& \int_0^1dx_1\int_0^1dx_2\,
    f_{W/e}(x_1)f_{W/e}(x_2)\,
    d\hat\sigma[W^+W^-\to W^+W^-; \hat s=x_1x_2s]
\end{eqnarray}
The c.m.\ invariant energy of the subprocess is given by $\sqrt{\hat
s}=\sqrt{x_1x_2s}$.  The fixing of final-state observables $\Omega$
can be implemented by restricting the integration over the phase space
$\hat\Phi$ appropriately:
\begin{equation}
  \frac{d\sigma}{d\Omega}[e^+e^-\to \bar\nu_e\nu_eW^+W^-]
  = \int_0^1dx_1\int_0^1dx_2\,
    f_{W/e}(x_1)f_{W/e}(x_2)\,
    \frac{d\hat\sigma}{d\hat\Phi}d\hat\Phi\,
    \delta\left(\Omega-\Omega(x_1,x_2)\right)
\end{equation}
Other $W,Z$ processes can be treated analogously.

The commonly used equivalent particle spectra
in the leading logarithmic approximation,
Eqs.(31)-(32), are derived in the small-angle limit with zero $p_\perp$.
To suppress background processes which are induced by
Weizs\"acker-Williams photons, it is necessary to consider 
the transverse momentum distribution of the $W$~boson pair.  
To high accuracy, the c.m.\ frame of $\gamma$-initiated subprocesses moves
parallel to the $e^\pm$ beams.
The $W$-initiated signal processes, by contrast, have transverse 
momenta of order $P_\perp(WW)\sim M_W$. 
Hence, the $\gamma$-initiated background processes can be
eliminated by cutting on the total transverse momentum of the
subprocess with respect to the $e^\pm$~beams. 
For the above reason, the usual leading logarithmic equivalent-particle 
approximation, (31)-(32), cannot be applied 
when a $P_\perp(WW)$ cut is imposed in the analysis.
In order to provide a guideline for the later more complete analysis,
we start with the improved equivalent-particle formalism\cite{wuki},
from which we derive the $P_\perp (WW)$ distribution. 
This can be most conveniently performed by relating
the $W$ transverse momentum to its virtual mass squared $q^2$:
\begin{equation}
  p_{\perp } = 
	\frac{\sqrt{s}}{2}(1-x)
	\sqrt{1-\left[1+\frac{2q^2}{s(1-x)}\right]^2}
\end{equation}
with the space-like $q^2$ bounded by $-s(1-x) \leq q^2 \leq 0$.
Expressed in terms of $q^2$, the improved equivalent particle
distributions can be written as
\begin{equation}\label{IEWA}
  f^{\lambda}_{W/e}(x) =
	\frac{\alpha x}{16\pi s_w}
	\int_{q^2_{\min}}^{q^2_{\max}}
	\frac{-q^2dq^2}{(q^2-M_W^2)^2}
	\left\{
	\begin{array}{ll}
	\displaystyle\frac{M_W^2\kappa_1^2}{-q^2}
		& \mbox{for $\lambda=L$}\\[0.45cm] 
	\displaystyle (1+\kappa_2^2)
		& \mbox{for $\lambda=T$}
	\end{array}   
	\right.
\end{equation}
where
\begin{equation}
  \kappa_1 \equiv
	\frac{2\sqrt{1-x+q^2/s}}{x-q^2/s} 
  \qquad
  \kappa_2 \equiv
	\frac{2}{x-q^2/s} - 1
\end{equation}
and $\lambda=L,T$ denotes longitudinal resp.\ transverse polarization.
In the latter case we have added the results for negative and positive
helicity of the $W$~boson.

The improved luminosity distributions of the $W$~bosons with respect to the
transverse momentum are thus given as follows:
\begin{equation}\label{IEWA-new}
  f^{\lambda}_{W/e}(x,p_\perp^2) =
	\frac{\alpha}{4\pi s_w^2}   
	\left\{
	\begin{array}{ll}
	\displaystyle\frac{\kappa_m x\kappa_1^2}
			{s\bar{x}r(\bar{x}\bar{r}+2\kappa_m)^2}  
		& ~~~~\mbox{for $\lambda=L$}\\[0.45cm] 
	\displaystyle\frac{x\bar{r}(1+\kappa_2^2)}
			{2sr(\bar{x}\bar{r}+2\kappa_m)^2}  
		& ~~~~\mbox{for $\lambda=T$}
	\end{array}
	\right.
\end{equation}
with
\begin{equation}
  \kappa_\perp\equiv \frac{{p_{\perp}}^2}{s}
  \qquad
  \kappa_m \equiv \frac{M_W^2}{s}
  \qquad
  r \equiv \sqrt{1-\frac{4\kappa_\perp}{(1-x)^2}}
  \qquad
  \bar{x}\equiv 1-x
  \qquad
  \bar{r}\equiv 1-r
\end{equation}

  In the asymptotic limit $s \gg P_\perp^2, M_W^2$, and for $x_{1,2}$
neither close to 0 nor 1, we can derive the following approximate formula
from eq.(\ref{IEWA-new}):
\begin{eqnarray}
  f^T_{W/e}(x,p_\perp^2) &=& \frac{\alpha}{4\pi s_w^2}\,
        \frac{1+(1-x)^2}{2x}\,
        \frac{p_\perp^2}{\left(p_\perp^2+(1-x)M_W^2\right)^2} 
  \label{fT-LLA}\\
  f^L_{W/e}(x,p_\perp^2) &=& \frac{\alpha}{4\pi s_w^2}\,
        \frac{1-x}{x}\,
        \frac{(1-x)M_W^2}{\left(p_\perp^2+(1-x)M_W^2\right)^2} 
  \label{fL-LLA}
\end{eqnarray}

The transverse momentum distribution 
($~f^{\lambda_1\lambda_2}_{WW/ee}(x,P_\perp^2 ;\hat{s} ) ~$)
of the two-particle $WW$ system can be approximately derived
by convoluting the spectra (\ref{IEWA-new}) for each initial $W$ boson:
\begin{eqnarray} 
\displaystyle 
d\sigma (e+e\rightarrow f_3+f_4+X\mid s)
=\sum_{\lambda}\int_0^1 dx \int dP_\perp^2
 f^{\lambda_1\lambda_2}_{WW/ee}(x,P_\perp^2 ;\hat{s} )
d\widehat{\sigma}(W_{\lambda_1}+W_{\lambda_2}\rightarrow X\mid\hat{s})\\[0.4cm]
   \label{fPTWW}
 f^{\lambda_1\lambda_2}_{WW/ee}(x,P_\perp^2 ;\hat{s} ) 
  = 	\int_x^1\!\int_0^1 dx_1dx_2
	\int\!\!\!\int d{p}_{1\perp}^2d{p}_{2\perp}^2
	D(x)\,D(P_\perp^2)\,
	f^{\lambda_1}_{W/e}(x_1,p_{1\perp}^2)\,
	f^{\lambda_2}_{W/e}(x_2,p_{2\perp}^2) 
\end{eqnarray}
with
\begin{eqnarray} 
\label{fPTWW2}
  D(x) &=&  \delta (x-x_1x_2)|_{\hat{s}=xs} \\
\label{fPTWW3}
  D(P_\perp^2 ) &=& 
	\int_0^{2\pi}\frac{d \varphi_{12}}{2\pi}\,
	\delta({P}_\perp^2 -|\vec{p}_{1\perp}+\vec{p}_{2\perp}|^2)
\end{eqnarray}
where $\varphi_{12}$ is the azimuthal angle between the two initial
$W$~bosons in the $e^+e^-$ c.m.\ frame.  Due to the implicit
$\varphi_{12}$-dependence in the squared transverse momentum,
$|\vec{p}_{1\perp}+\vec{p}_{2\perp}|^2$, the integral in (\ref{fPTWW3})
is non-trivial.

\begin{figure}
\vspace*{-1.5cm}
\begin{center}
\includegraphics[width=17cm,height=15cm]{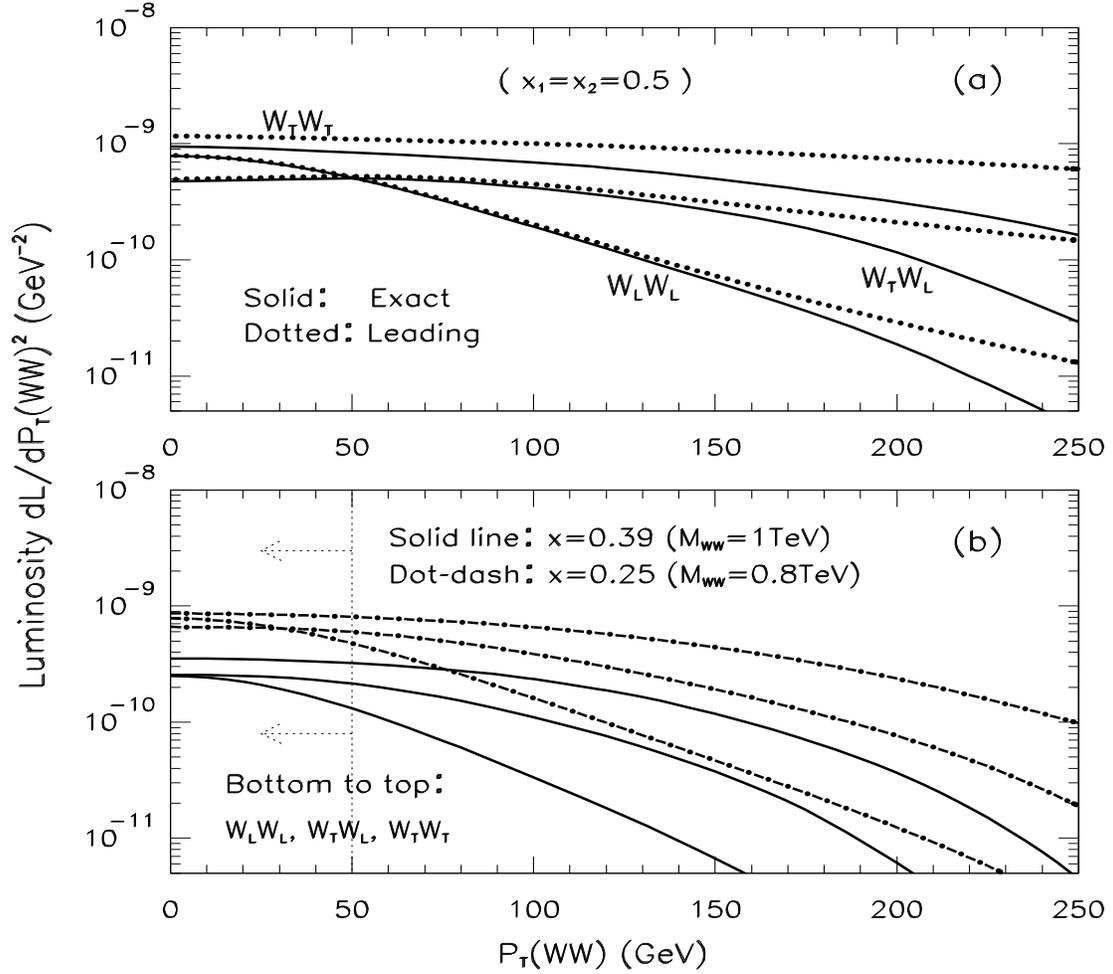}
\vspace*{-1.4cm}
\caption{Distribution of the $WW$ transverse momentum  $P_\perp (WW)$ 
         in $1.6\;{\rm TeV}$ $e^\pm e^-$ collisions.}
\label{fig:EWA-pt}
\end{center}
\end{figure}

The characteristic features of the luminosity spectra with respect to
the transverse momentum of the $WW$~system are exemplified in
Fig.\ref{fig:EWA-pt}.  In Fig.\ref{fig:EWA-pt}(a) we depict the $P_\perp
(WW)$ distributions for $x_1=x_2=0.5$.  The probability for the
emission of longitudinal $W$ bosons is maximal around low values of
$P_\perp\sim M_W/2$ and falls off rapidly with increasing transverse
momentum. The spectrum of the transverse $W$ bosons extends to much
larger values of $P_\perp$, decreasing asymptotically like
$(1/P_\perp^2)\ln(M_W^2/P_\perp^2)$.  The $\gamma\gamma$ spectrum, by
contrast, is strongly peaked at zero transverse momentum.

Since the phase space in~(\ref{fPTWW}) is restricted for a finite
collider energy~$\sqrt{s}$, the improved distributions (\ref{IEWA-new})
[solid lines in Fig.\ref{fig:EWA-pt}(a)] decrease for large transverse
momenta faster than the approximate distributions
(\ref{fT-LLA}-\ref{fL-LLA}) [dotted].

In Fig.\ref{fig:EWA-pt}(b) the $WW$ transverse momentum distributions
(\ref{fPTWW}) are depicted for two values of the invariant $WW$ mass,
$M_{WW}=0.8$ and $1\;{\rm TeV}$, at a fixed collider energy of
$\sqrt{s}=1.6\;{\rm TeV}$.  A typical cut of $50\;{\rm GeV}$, which
will be introduced below (cf.\ Sec.\ref{sec:comphep}), is indicated by
the dotted line.  The distributions are not shown for transverse
momenta beyond $\sim 250\;{\rm GeV}$ since interference effects
between the amplitudes become significant for large transverse
momenta, invalidating the probabilistic picture of the single-particle
distributions.

As shown in Fig.\ref{fig:EWA-pt}(b), 
for higher values of $M_{WW}$ the distributions are shifted 
towards lower values of $P_\perp^2$. 
For a fixed $M_{WW}$, the improved $P_\perp$ distributions
are lower than the approximate ones, as shown in Fig.\ref{fig:EWA-pt}(a). 
For this reason, the leading logarithmic approximation generally
overestimates the production rates due to 
transverse $W$~boson fusion by a factor of $3\sim 5$~\cite{LLEWA-VT,kuss-VT}.  
Therefore, we use the improved equivalent-particle method, 
in contrast to the leading logarithmic approximation, 
as a guideline for the analysis and as an
independent check for the complete tree-level calculation.
It turns out that the $\pm 1\sigma$ exclusion contours for $\alpha_{4-10}$, 
as shown in Figs.\ref{contour1600}, \ref{contour67} and \ref{alpha10},
obtained from the above two methods, are  in good agreement
after imposing all the relevant  kinematic cuts to enhance the ratio of 
signal to background.\footnote{
For example, the $90\%$ exclusion limit, obtained from using the 
improved equivalent-particle method which predicts
a nontrivial $P_\perp (WW)$ distribution, agrees with that in 
Figs.\ref{contour1600} at the level of $20$-$30\%$.}
Hence, we shall not discuss in detail the numerical results obtained 
by applying the equivalent-particle method, but we will focus on the 
improved results which are based on the exact tree-level calculations.

\section{Calculation and results: Conserved custodial \boldmath$SU(2)_c$}
\label{sec:comphep}
For a detailed numerical study, based on a complete tree-level
calculation, we have chosen the three processes
\begin{eqnarray}
  e^+e^- &\to& \bar\nu_e\nu_eW^+W^-\label{WWnn}\\
  e^+e^- &\to& \bar\nu_e\nu_eZZ    \label{ZZnn}\\
  e^-e^- &\to& \nu_e\nu_eW^-W^-    \label{sWWnn}
\end{eqnarray}
where the (quasi-)elastic $WW$ scattering signal
corresponds to the generic diagrams depicted in
Fig.\ref{graphs:signal}.
However, there are also Feynman diagrams contributing to
(\ref{WWnn}--\ref{sWWnn}) which do not contain $WW$ scattering
as a subprocess (cf.~Fig.\ref{graphs:irr-bg}).  This irreducible
background is not negligible and must be taken into account in 
the analysis.

In all signal processes there are already two neutrinos present
in the final state, therefore
important kinematic information is lost if a
$W$~boson decays leptonically (or a $Z$ boson into two neutrinos).  In
particular, the c.m.\ energy of the subprocess cannot be determined in
that case.  For the present study we therefore restrict ourselves to
hadronic $W,Z$ decays and to decays of the $Z$ boson into electrons
and muons.  Furthermore, an error in the dijet invariant mass is
introduced by the limited energy resolution of the calorimeters, which
leads to the rejection of a fraction of di-boson events and to the
misidentification of $W$ \emph{vs.}\ $Z$ bosons.  Adopting the results
for the net efficiencies determined in Ref.\cite{BCHP}, we assume that
in a hadronic decay a true $W,Z$ boson will be identified according to
the following pattern\footnote{Using the tagging of $b$-quarks, the $Z\to W$
misidentification probability could be reduced, thus improving
its detection efficiency.}:
\begin{eqnarray}
  \label{W-eff}
  W &\to& 85\%\;W,\ 10\%\;Z,\ 5\%\;\mbox{reject}\\
  \label{Z-eff}
  Z &\to& 22\%\;W,\ 74\%\;Z,\ 4\%\;\mbox{reject}
\end{eqnarray}
Thus, for example, when calculating the signal event rate in
the $ZZ$ detection mode, 
one has to include the rates predicted by 
55\%, 7\%, and 1\% of the partonic $ZZ$, $W^\pm Z$ and $W^+W^-$ 
final states, respectively, to account for final-state misidentification.
The relative weighting factors from the above three partonic final state
cross sections are $55 : 7 : 1$ which is equal to $1: 0.13 : 0.018$, 
as given in the last column of Tab.II.
As discussed above, we only consider the hadronic decay modes of 
a final state $W$ boson, the corresponding decay branching ratio (BR) 
is $0.67$. For detecting a $Z$ boson, we include both the
hadronic modes (BR=0.70) and the di-lepton modes (BR=0.067 for
 $e^+e^-$ and $\mu^+\mu^-$). 
Hence, the efficiency for detecting a $WW$, $ZZ$, and $WZ$ pair
originating from a partonic $WW$, $ZZ$, and $WZ$ final state is
33.4\%, 34.2\%, and 33.8\%, respectively. For simplicity,
we take 33\% as the detection efficiency for all the
detection modes considered in this study.

\newcommand{\eegraph}[2]{%
\begin{fmfgraph*}(40,30)
  \fmfleftn{i}{6} \fmfrightn{o}{4}
  \fmf{fermion,tens=#2}{i2,v1} \fmf{fermion,tens=#2}{v2,i5}
  #1
\end{fmfgraph*}%
}
\newcommand{\eewwgraph}[1]{%
\eegraph{\fmf{fermion}{v1,o1} \fmf{fermion}{o4,v2}
         \fmf{phantom}{v1,v3,o2} \fmf{phantom}{o3,v4,v2}
         \fmf{phantom}{v4,v3}\fmffreeze #1}{2}}

\begin{figure}
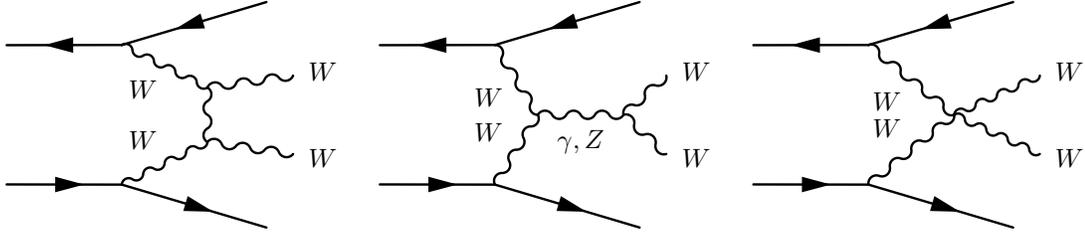

\begin{center}
\unitlength1mm
\eewwgraph{\fmf{boson,lab=\fn$W$}{v1,v5}\fmf{boson}{v5,o2} 
           \fmf{boson}{o3,v6}\fmf{boson,lab=\fn$W$}{v6,v2} 
           \fmflabel{\fn$W$}{o2}\fmflabel{\fn$W$}{o3}
           \fmf{boson}{v6,v5}}
\qquad
\eewwgraph{\fmf{boson,lab=\fn$W$}{v1,v5}
           \fmf{boson,lab=\fn$W$,lab.s=left}{v5,v2} 
           \fmf{boson}{o2,v6}\fmf{boson}{v6,o3} 
           \fmflabel{\fn$W$}{o2}\fmflabel{\fn$W$}{o3}
           \fmf{boson,lab=\fn$\gamma,,Z$,lab.s=left}{v6,v5}}
\qquad
\eewwgraph{\fmf{boson,lab=\fn$W$}{v1,v5}\fmf{boson}{v5,o2} 
           \fmf{boson}{o3,v5}\fmf{boson,lab=\fn$W$,lab.s=left}{v5,v2}
           \fmflabel{\fn$W$}{o2}\fmflabel{\fn$W$}{o3}}
\end{center}
\caption{Diagrams contributing to the strong $WW$ scattering signal.}
\label{graphs:signal}
\end{figure}
\begin{figure}
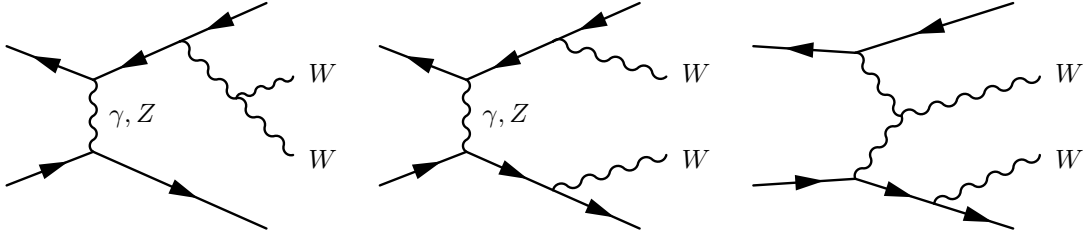

\begin{center}
\unitlength1mm
\eegraph{\fmf{phantom}{v1,v4,o1} \fmf{fermion}{o4,v3,v2} 
         \fmf{boson,lab=\fn$\gamma,,Z$}{v1,v2}
         \fmffreeze \fmf{fermion}{v1,o1}
         \fmf{boson}{v3,v5,o2} \fmffreeze \fmf{boson}{v5,o3}
         \fmflabel{\fn$W$}{o3}\fmflabel{\fn$W$}{o2}}{1}
\qquad
\eegraph{\fmf{fermion}{v1,v3,o1} \fmf{fermion}{o4,v4,v2}
         \fmf{boson,lab=\fn$\gamma,,Z$}{v1,v2}
         \fmffreeze \fmf{boson}{v3,o2} \fmf{boson}{o3,v4}
         \fmflabel{\fn$W$}{o3}\fmflabel{\fn$W$}{o2}}{1}
\qquad
\eegraph{\fmf{phantom}{v1,o1}\fmf{fermion}{o4,v2} 
         \fmf{boson}{v1,v3,v2} \fmf{phantom,tens=1/3}{o2,v3,o3}
         \fmffreeze \fmf{boson}{o3,v3}
         \fmf{fermion}{v1,v4,o1}\fmffreeze\fmf{boson}{o2,v4}
         \fmflabel{\fn$W$}{o3}\fmflabel{\fn$W$}{o2}}{2}
\caption{Typical diagrams contributing to the irreducible background
for the strong $WW$ scattering signal.}
\label{graphs:irr-bg}
\end{center}
\end{figure}
\begin{figure}
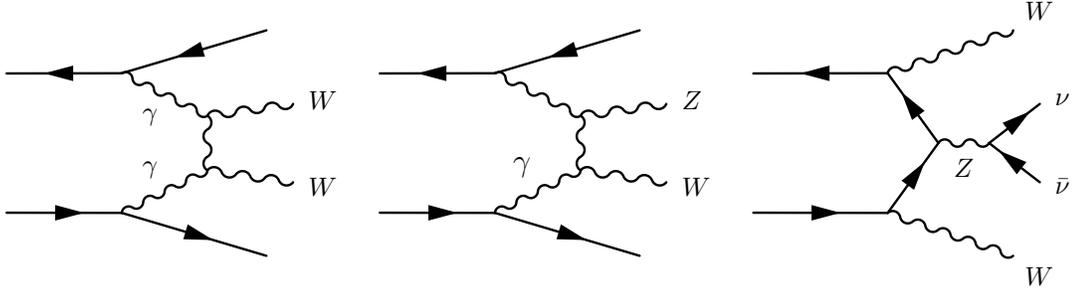

\begin{center}
\unitlength1mm
\eewwgraph{\fmf{boson,lab=\fn$\gamma$}{v1,v5} \fmf{boson}{v5,o2} 
           \fmf{boson,lab=\fn$\gamma$}{v6,v2} \fmf{boson}{o3,v6}
           \fmf{boson}{v6,v5}  
           \fmflabel{\fn$W$}{o3} \fmflabel{\fn$W$}{o2} 
}
\qquad
\eewwgraph{\fmf{boson,lab=$\gamma$}{v1,v5} \fmf{boson}{v5,o2} 
           \fmf{boson}{o3,v6,v2} \fmf{boson}{v6,v5}  
           \fmflabel{\fn$Z$}{o3} \fmflabel{\fn$W$}{o2} 
}
\qquad
\eegraph{\fmf{boson}{v1,o1}  \fmf{boson}{o4,v2}
         \fmf{phantom}{v1,v3,o2} \fmf{phantom}{o3,v4,v2}
         \fmf{phantom}{v4,v3}\fmffreeze
         \fmf{fermion}{v1,v5,v2}
         \fmf{boson,tens=2,lab=\fn$Z$}{v5,v6}  \fmf{fermion}{o2,v6,o3}
         \fmflabel{\fn$\bar\nu$}{o2}  \fmflabel{\fn$\nu$}{o3}
         \fmflabel{\fn$W$}{o1}\fmflabel{\fn$W$}{o4}
         }{1.5}
\end{center}
\caption{Partially reducible backgrounds to the strong $WW$
scattering signal.}
\label{graphs:red-bg}
\end{figure}

Since the final state cannot be completely resolved experimentally in
all cases, further background processes will play a role (cf.\
Fig.\ref{graphs:red-bg}).  The most important background to the signal
process $e^+e^-\to \bar\nu\nu W^+W^-$ is generated by the reaction
\begin{equation}
  e^+e^- \to W^+W^- e^+e^-
\end{equation}
which is built up primarily by the subprocess $\gamma\gamma\to
W^+W^-$.  In this process most of the electrons/positrons are emitted
in forward direction so that they cannot be detected.  A similar
background is introduced by the misidentification of vector bosons in
jet decays:
\begin{equation}
  e^+e^- \to W^\pm Z e^\mp \nu
\end{equation}
An irreducible background is also generated by three-boson final states,
\begin{equation}
  e^+e^- \to W^+W^-Z
\end{equation}
with the $Z$ decaying into neutrino pairs.  Similar backgrounds (less
dangerous for the $ZZ$ final state) exist for the other processes.

The total cross sections for the signal and background processes,
including interference effects, have been computed in a complete
tree-level calculation using the automatic package {\tt
CompHEP}~\cite{CompHEP} in which the effective Lagrangian~(\ref{chL})
has been implemented.  The results of the cross sections for the
reference point $\alpha_4=\alpha_5=0$ are summarized in
Tab.\ref{tab:nocuts}.

\begin{table}
\begin{displaymath}
\begin{array}{|l||r|r||c|}
\hline
\mbox{Process} & {800}\ {\rm GeV} & 
{1.6}\ {\rm TeV} & \mbox{Factor}\\
\hline\hline
W^+W^-\bar\nu\nu        &  11 &   56 & 1\\
W^+W^-e^+e^-            & 628 & 1979 & 1\\
W^\pm Z e^\mp\nu        &  39 &  173 & 0.26\\
W^+W^-(Z\to\bar\nu\nu)  &  13 &   11 & 1\\
\hline\hline
ZZ\bar\nu\nu            &   4 &   26 & 1 \\
ZZe^+e^-                &   2 &    4 & 1 \\
W^\pm Z e^\mp\nu        &  39 &  173 & 0.13\\
W^+W^-e^+e^-            & 628 & 1979 & 0.018\\
ZZ(Z\to\bar\nu\nu)      & 0.6 &  0.4 & 1\\
\hline\hline
W^-W^-\nu\nu            &  14 &   67 & 1\\
\hline
\end{array}
\end{displaymath}
\caption{Total cross sections in $\rm fb$ for various processes.
Dectection efficiencies and branching ratios are not included.
Including final-state misidentification, the numbers should be
multiplied by the relative weighting
factor given in the last column which accounts for
final-state misidentification in the corresponding detection mode
($W^+W^-$, $ZZ$, or $W^-W^-$).}
\label{tab:nocuts}
\end{table}

The background reduction is essential for isolating the strong
scattering signal, as demonstrated by the numbers in
Tab.\ref{tab:nocuts}. To this purpose, we follow the strategy introduced
in Ref.\cite{BCHP}:
\begin{enumerate}
\item We require $M_{\rm inv}(\bar\nu\nu)>200\,(150)\ {\rm GeV}$.  The
first number applies for $\sqrt{s}=1.6\;{\rm TeV}$, while the
bracketed number for $\sqrt{s}=800\;{\rm GeV}$.  This cut removes the
events with neutrinos from $Z$ decay together with backgrounds from
$W^+W^-$ and QCD four-jet production.  The signal is not affected
(cf.~Fig.\ref{mNN}).
\begin{figure}
\includegraphics{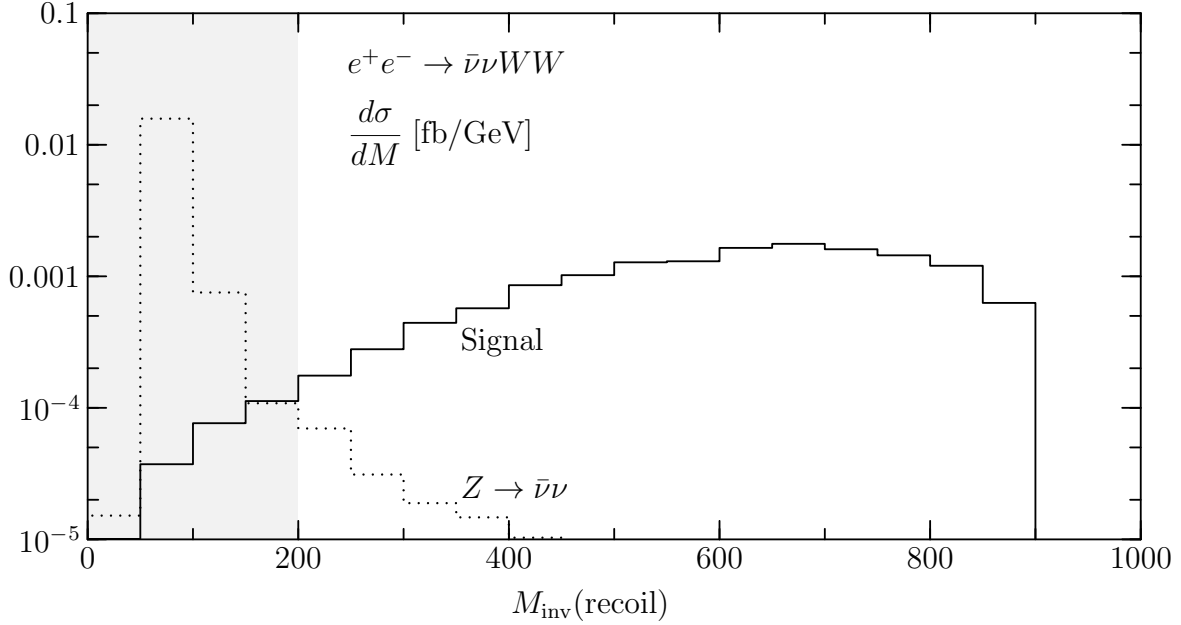}
\caption{Distribution of the invariant $WW$ recoil mass distribution
in the process $e^+e^-\to W^+W^-\bar\nu\nu$ (signal).  The cut (shaded
area) removes events in which the neutrinos are generated through $Z$
decays.  The other cuts have been applied as described in the text.}
\label{mNN} 
\end{figure}
\item Selecting central events [$|\cos\theta(W/Z)|<0.8$] with
$p_\perp(W/Z)>200\,(100)\ {\rm GeV}$ removes events dominated by
$t$-channel exchange in the subprocess.
\item The background from $\gamma\gamma$ fusion is reduced by two
orders of magnitude if an electron veto is applied [removing events
with $\theta(e)>10^\circ$] and, at the same time, a minimum $p_\perp$
of the vector boson pair, equivalent to the fermion $p_\perp$, is
required.  We use $p_\perp(WW)>50\,(40)\ {\rm GeV}$ and
$p_\perp(ZZ)>30\ {\rm GeV}$.  This cut removes also about one half of
the signal events.  (Fig.\ref{pt2W}; cf.\ the discussion in
Sec.\ref{sec:IEWA})
\enlargethispage{2mm}
\item Since the impact of the strong interaction terms ${\cal L}_4$
and ${\cal L}_5$ increases with the energy of the subprocess, we use a
window in $M_{\rm inv}(WW/ZZ)$ between $700\,(350)$ and $1200\,(600)\
{\rm GeV}$, Fig.\ref{mWW}.  [The bulk of events has lower invariant
mass, but those are quite insensitive to the parameters of interest.]
\end{enumerate}
\begin{figure}[p]
\includegraphics{wwdist.3}
\vskip-2mm
\caption{Transverse momentum distribution of the $W$ pair in the
process $e^+e^-\to W^+W^-\bar\nu\nu$ at $\protect\sqrt{s}=1.6\;{\rm
TeV}$.  All cuts have been applied, but the $WW$ detection efficiency
(therefore, the decay branching ratio)
is not included.  The solid line
corresponds to the reference point $\alpha_4=\alpha_5=0$, the dashed
line to $\alpha_4=0.005$ for comparison.  The dominant backgrounds
$W^+W^-e^+e^-$ (dotted) and $WZe\nu$ (dot-dashed, with $26\%$
misidentification probability) are also indicated.  The shaded area is
removed by the $p_\perp$ cut.}
\label{pt2W}
\vspace*{8mm}
\includegraphics{wwdist.2}
\vskip-2mm
\caption{Invariant mass distribution of the $W$ pair in the process
$e^+e^-\to W^+W^-\bar\nu\nu$.  Legend as in Fig.\ref{pt2W}.}
\label{mWW}
\end{figure}
After applying those cuts, we find the numbers reported in
Tab.\ref{tab:allcuts}.  If they are multiplied by the
misidentification probabilities in the last column, the
signal/background ratios are raised to~$O(1)$.
\begin{table}
\begin{displaymath}
\begin{array}{|l||r|r||c|}
\hline
\mbox{Process} & {800}\ {\rm GeV} & 
{1.6}\ {\rm TeV} & \mbox{Factor}\\
\hline\hline
W^+W^-\bar\nu\nu        & 0.41 & 0.71 & 1\\
W^+W^-e^+e^-            & 0.12 & 0.47 & 1\\
W^\pm Z e^\mp\nu        & 1.42 & 1.23 & 0.26\\
W^+W^-(Z\to\bar\nu\nu)  & 0.01 & 0.01 & 1\\
\hline\hline
ZZ\bar\nu\nu            & 0.33 & 0.86 & 1 \\
ZZe^+e^-                & 0.00 & 0.00 & 1 \\
W^\pm Z e^\mp\nu        & 1.54 & 1.37 & 0.13\\
W^+W^-e^+e^-            & 0.51 & 0.93 & 0.018\\
ZZ(Z\to\bar\nu\nu)      & 0.00 & 0.00 & 1\\
\hline\hline
W^-W^-\nu\nu            & 0.81 & 1.36 & 1\\
\hline
\end{array}
\end{displaymath}
\caption{Cross sections in $\rm fb$ as in Tab.\ref{tab:nocuts}, but
including all cuts.}
\label{tab:allcuts}
\end{table}
In order to obtain the final event rates, the cross sections in
Tab.\ref{tab:allcuts} have to be multiplied by the expected luminosity
and $33\%$ detection efficiency [cf.(\ref{W-eff}-\ref{Z-eff}); this
number includes the $W/Z$ decay branching ratios].

For polarized beams with left-handed electron and right-handed
positron polarizations $P_\mp$, the rates are modified as follows:
\begin{enumerate}
\item Two left-handed electron/positron couplings are involved in the
signal process.  The rate is therefore increased by the factor
$(1+P_+)(1+P_-)$.
\item The dominant part of the $WZ$ background is initiated by $\gamma
W$ fusion which involves only one left-handed coupling.  The cross
section is therefore increased by the factor $1 + (P_+ + P_-)/2$.
Since the $Z$ coupling to electrons is almost of axial-vector type,
this holds approximately true also for the remainder of the $WZ$
background.
\item The $WWee$ background is not modified.  [There are diagrams in
which the $W$'s both originate from the same fermion line.  The
contribution from this kind of diagrams should increase by the factor
$1 + (P_+ + P_-)/2$; however, its net effect is not important.]
\end{enumerate}
We conclude that \emph{both} electron and positron polarization is
essential in order to improve the signal rate as well as the
signal/background ratio.  In the ideal case of complete polarization,
$S/B$ improves by a factor~$2$ and $S/\sqrt{B}$ by a factor~$3$ as far
as reducible backgrounds are concerned.  For the irreducible part,
$S/\sqrt{B}$ increases by a factor~$2$ from the rate alone.

\begin{figure}
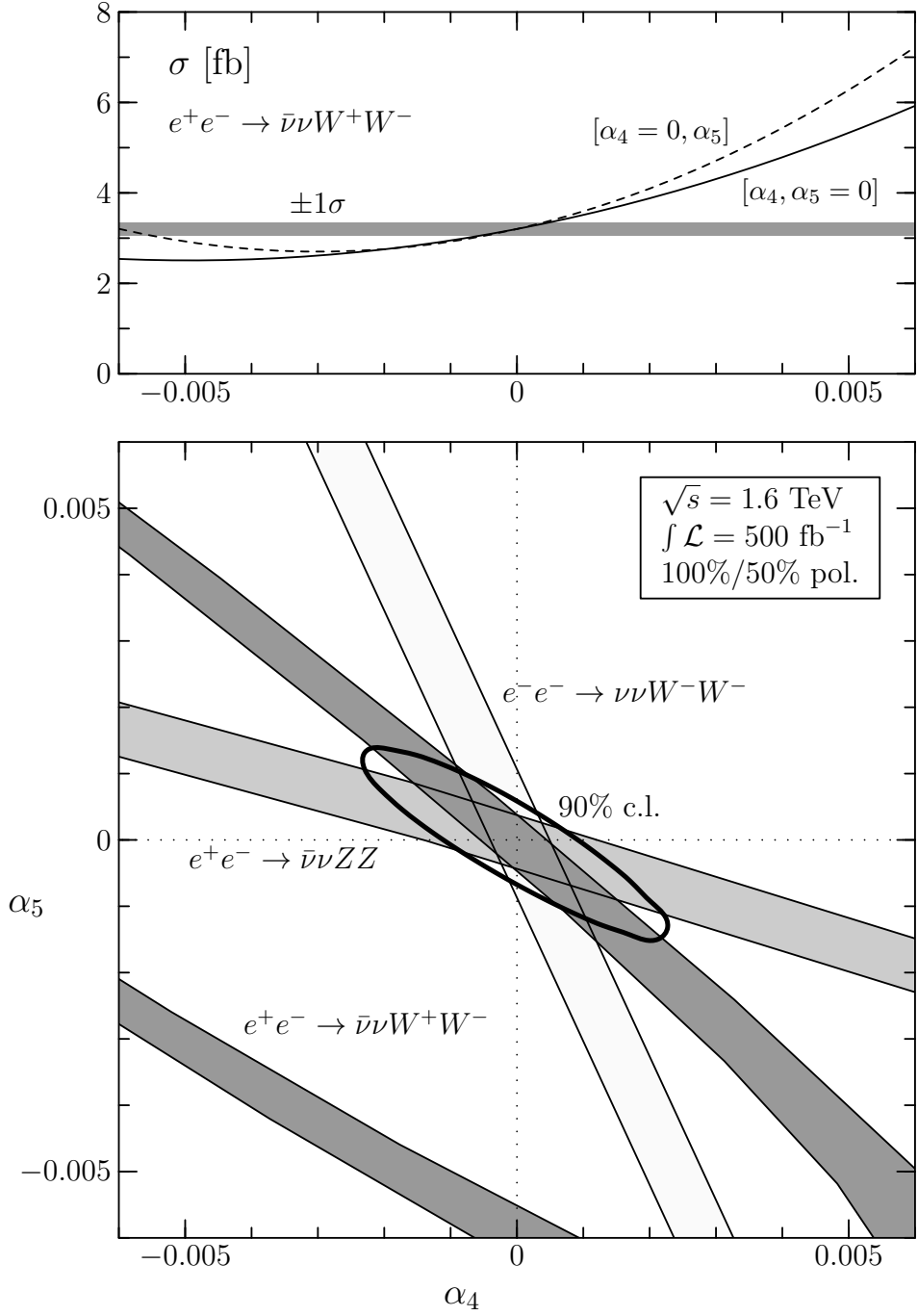

\begin{center}
\unitlength1mm
\includegraphics{contour.1}\\[9mm]
\includegraphics{contour.2}\\[10mm]
\end{center}
\caption{Upper part: Cross section including backgrounds for the
process $e^+e^-\to \bar\nu\nu W^+W^-$.  All cuts have been applied.
The shaded band is the statistical error corresponding to the expected
detection efficiency and a luminosity of $\int{\cal L}=500\;{\rm
fb}^{-1}$.  It is assumed that the $e^-$($e^+$) beam is polarized at a
degree of $100\%\;(50\%)$.  Lower Part: $1\sigma$ exclusion contours
for all three processes in the $\alpha_4/\alpha_5$ plane, based on the
hypothesis $\alpha_4=\alpha_5=0$.  All cuts have been applied and
detection efficiencies are included.  The closed contour curve is the
90\%\ exclusion limit obtained by combining the $e^+e^-\to \bar\nu\nu
W^+W^-$ and $e^+e^-\to \bar\nu\nu ZZ$ channels.}
\label{contour1600}
\end{figure}

All numbers quoted so far were based on the values
$\alpha_4=\alpha_5=0$.  Ultimately we are interested in the
measurement of those parameters.  The result of the theoretical
prediction is depicted in Fig.\ref{contour1600}.  In the upper part
the dependence of the cross sections on $\alpha_4$ and $\alpha_5$ is
displayed for polarized beams after all cuts are applied, but no
detection efficiencies included.  The band, based on the hypothesis
$\alpha_4=\alpha_5=0$, is determined by the $\pm 1\sigma$ statistical
error in the $WW\bar\nu\nu$ event rate if the expected integrated
luminosity of $\int{\cal L}=500\;{\rm fb}^{-1}$ and the efficiency of
$33\%$ [which includes the $W/Z$ decay branching ratios] are taken
into account.  The lower part of the figures shows the corresponding
experimental regions in the two-dimensional $[\alpha_4,\alpha_5]$
plane, based on the hypothesis $\alpha_4=\alpha_5=0$.  We display the
($\pm 1\sigma$) bounds for the individual channels, which can be
combined to give the $90\%$ exclusion limit indicated by the closed
contour curve.

\section{Calculation and results: Broken custodial \boldmath$SU(2)_c$}
\label{sec:SU2c}
In addition to the interactions {\small ${\cal L}_{4,5}$} in
(\ref{L4}--\ref{L5}), three more dimension-$4$ operators ${\cal
L}_{6,7,10}$ are present at next-to-leading order of the electroweak
chiral Lagrangian.  Since these interactions affect the quartic gauge
couplings only, they also do not contribute to low-energy observables
at tree level:
\begin{eqnarray}
  {\cal L}_6 &=&
	\alpha_6\,{\rm tr}\left[V_\mu V_\nu\right]
	{\rm tr}\left[{\cal T}V^\mu\right]
	{\rm tr}\left[{\cal T}V^\nu\right]
	\\
  {\cal L}_7 &=&
	\alpha_7\,{\rm tr}\left[V_\mu V^\mu\right]
	{\rm tr}\left[{\cal T}V_\nu\right]
	{\rm tr}\left[{\cal T}V^\nu\right]
	\\
  {\cal L}_{10} &=&
	\alpha_{10}\,
	\frac{1}{2}\left(
	{\rm tr}\left[{\cal T}V^\mu\right]
	{\rm tr}\left[{\cal T}V^\nu\right]\right)^2
\end{eqnarray}
where ${\cal T}=U\tau_3 U^\dagger$.  Due to the presence of ${\cal
T}$, the new operators ${\cal L}_{6,7,10}$ violate the custodial
$SU(2)_c$ symmetry in contrast to ${\cal L}_{4,5}$.

The coefficients $\alpha_{4,5}$ and $\alpha_{6,7,10}$ can be
constrained only indirectly from low-energy observables, to which they
contribute through one-loop diagrams at the order of 
$\alpha_n {1 \over 16 \pi^2} \sim {v^2\over \Lambda^2}
{1 \over 16 \pi^2}$~\cite{HKY97}\footnote{
Here, $\Lambda \lesssim 4\pi v \sim 3$~TeV\cite{NDA} is the cut-off
of the effective Lagrangean, which characterizes the scale of the 
new strong interactions.}.   Since the corresponding loop
divergences must be absorbed by renormalization counterterms, it is
impossible to derive \emph{precise} bounds on these parameters from
low-energy data.  Nevertheless, rough estimates can be obtained by
keeping only the leading logarithmic terms.  The estimated indirect
bounds on these $4$-boson couplings are summarized in the following
list~\cite{lowbound,HKY97}
\begin{equation}  \label{bound}
\begin{array}{c@{\qquad}c}
-25\times 10^{-3} \leq \alpha_4 \leq 125 \times 10^{-3}&
-4 \times 10^{-3} \leq \alpha_6 \leq 22  \times 10^{-3}\\
-63\times 10^{-3} \leq \alpha_5 \leq 318 \times 10^{-3}&   
-32\times 10^{-3} \leq \alpha_7 \leq 163 \times 10^{-3}\\
&  
-4\times 10^{-3} \leq \alpha_{10} \leq 22\times 10^{-3}
\end{array}
\end{equation}
which are derived at $90\%$ c.l.\ by setting only one new parameter
nonzero at a time.  Even though current bounds on the $\rho$~parameter
severely constrain the possible amount of $SU(2)_c$ violation, the
next-to-leading $SU(2)_c$-violating parameters $\alpha_{6,7,10}$ are
still allowed in the range from $0.02$ to $0.2$ which is well above
the natural value $\sim 1/16\pi^2\simeq 0.006$.

In this section, we focus on tests of the $SU(2)_c$-violating
operators ${\cal L}_{6,7,10}$ in quasi-elastic $WW$ scattering.
Unlike the parameters $\alpha_{4,5}$, the terms $\alpha_{6,7,10}$
signal new dynamics beyond the standard model (SM), since the SM-like
Higgs sector respects $SU(2)_c$-symmetry and thus does not contribute
to $\alpha_{6,7,10}$.  The leading contribution of the quasi-elastic $WW\to
WW$ scattering amplitudes is associated with longitudinal gauge bosons
and can be written as follows:
\begin{eqnarray}
  A(W^+W^-\to W^+W^-) 
       &=&  -\frac{u}{v^2} +
       \frac{4(s^2+t^2+2u^2)}{v^4}\alpha_4
            +\frac{8(s^2+t^2)}{v^4}\alpha_5  \label{4W-1} \\[0.25cm]
  A(W^+W^-\to ZZ) 
       &=&  +\frac{s}{v^2} +
       \frac{4(t^2+u^2)}{v^4}(\alpha_{4}+\alpha_{6})
       + \frac{8s^2}{v^4}(\alpha_{5}+\alpha_{7}) \label{4W-2}   \\[0.25cm]
  A(W^- W^- \to W^- W^- ) 
       &=&  -\frac{s}{v^2} + 
       \frac{4(2s^2+t^2+u^2)}{v^4}\alpha_4
       +\frac{8(t^2+u^2)}{v^4}\alpha_5  \label{4W-3} \\[0.25cm]
  A(W^\pm Z\to W^\pm Z) 
       &=&  +\frac{t}{v^2} +
       \frac{4(s^2+u^2)}{v^4}(\alpha_{4}+\alpha_{6})
       + \frac{8t^2}{v^4}(\alpha_{5}+\alpha_{7}) \label{4W-4}  \\[0.25cm]
  A(ZZ\to ZZ) 
       &=&   0 +  
       \frac{8(s^2+t^2+u^2)}{v^4}
       \left[(\alpha_{4}+\alpha_{5})
       +2(\alpha_{6}+\alpha_{7}+\alpha_{10})
       \right] \label{4W-5} 
\end{eqnarray}
The amplitudes are given for asymptotic energies at which the $W,Z$
masses can be neglected.

The five parameters $\{\alpha_{4,5};\;\alpha_{6,7,10}\}$ can in
principle be uniquely determined by measuring the total cross sections
of the processes (\ref{4W-1}--\ref{4W-5}).  If the event rates are
large enough, additional information can be extracted from the
$M_{WW}$, $P_{\perp}$, and angular distributions.  However, due to
large backgrounds and the small $eeZ$ coupling, the experimental
analysis of the reactions (\ref{4W-4}--\ref{4W-5}) is more difficult.

Elastic $W^-W^+\to W^-W^+$ and $W^-W^-\to W^-W^-$ scattering depends
only on $\alpha_{4}$ and $\alpha_{5}$; these two processes are
sufficient to determine both $\alpha_{4}$ and $\alpha_{5}$ to a high
accuracy (Fig.\ref{contour1600}).  The two reactions can therefore be
taken as reference processes.  The other two processes $W^-W^+\to ZZ$
and $W^\mp Z\to W^\mp Z$ can subsequently be exploited to measure
$\alpha_6$ and $\alpha_7$, while $\alpha_{10}$ can finally be
extracted from the reaction $ZZ\to ZZ$.

\begin{figure}
\begin{center}
\includegraphics{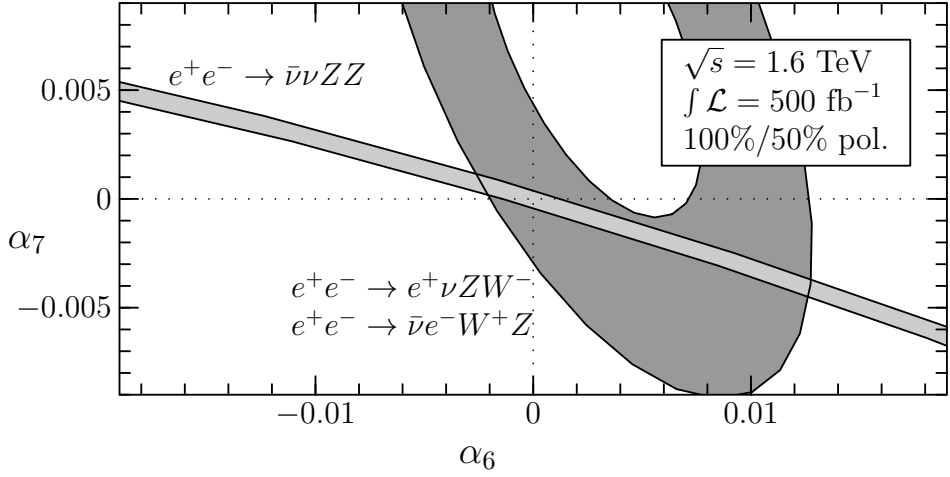}
\vskip15mm
\caption{$1\sigma$ exclusion contours for the
$SU(2)_c$-violating parameters $\alpha_{6,7}$ from $e^-e^+\to
\nu\bar{\nu} ZZ$ and $e^-e^+\to e^-\bar{\nu} W^+Z/e^+{\nu} W^-Z$.
All cuts have been applied as described in the text, and the detection
effiencies are included.}
\label{contour67}
\end{center}
\end{figure}

\begin{figure}
\begin{center}
\includegraphics{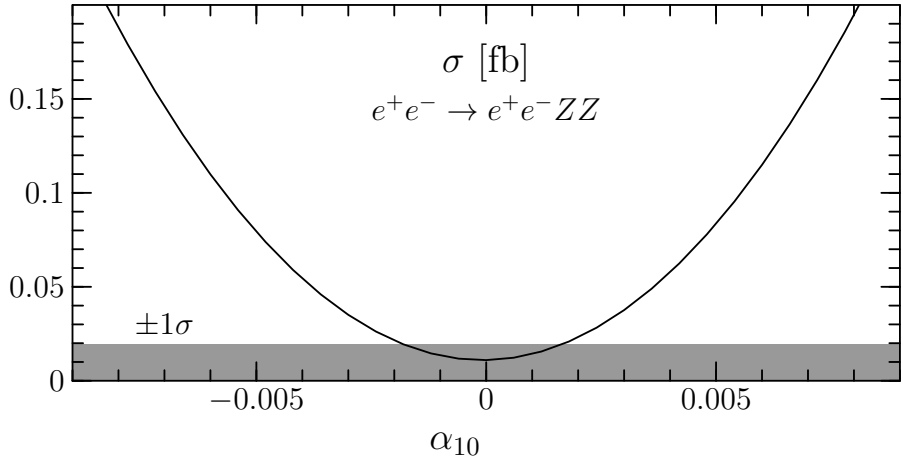}
\vskip15mm
\caption{Cross section (including backgrounds and cuts) for
$e^-e^+\to e^-e^+ ZZ$ as a probe of $\alpha_{10}$.  The shaded band is
the statistical error corresponding to the expected detection
efficiency and an integrated luminosity of $500\;{\rm fb}^{-1}$.}
\label{alpha10}
\end{center}
\end{figure}

To probe the chiral parameters $\alpha_{6}$, $\alpha_{7}$, and
$\alpha_{10}$, we assume that the $SU(2)_c$-conserving parameters
$\alpha_4$ and $\alpha_5$ have been pre-determined in the processes
$e^+e^-\to \bar\nu\nu W^+W^-$ and $e^-e^-\to\nu\nu W^-W^-$;\footnote{
As indicated in Fig.\ref{contour1600}, 
measuring the event rates of these two processes only, in general leads
to two allowed regions in the $\{\alpha_4,~\alpha_5\}$ plane. 
They can in principle be separated by carefully studying 
various distributions, which is beyond the scope of the present work.}~ 
in the 
following analysis we therefore set these parameters to the reference
values $\{0,0\}$ \emph{sine restructione generalitis}.  In this framework,
the $\pm 1\sigma$ exclusion contours for $\alpha_6$ and $\alpha_7$ are
shown in Fig.\ref{contour67} for the reactions $e^+e^-\to e^+\nu
W^-Z+\mbox{c.c.}$ and $e^+e^-\to\bar\nu\nu ZZ$.  The $e\nu WZ$ final
states suffer from large backgrounds due to $\gamma$-induced $eeWW$
events in which one $e$ is lost in the beampipe and one $W$
misidentified as $Z$.  This background can be suppressed efficiently
by a cut in the missing transverse momentum which in the following
analysis is set to $p_\perp({\rm miss.})>30\;{\rm GeV}$.  To isolate
the signal, we furthermore require the final-state electron to be
detected ($\theta>10^\circ$) and apply the additional cuts described
in Sec.\ref{sec:comphep}, with the exception of the cut on the
boson pair transverse momentum which is not useful here.  

The remaining chiral parameter $\alpha_{10}$ can be determined in the
process $e^+e^-\to e^+e^- ZZ$.  Since elastic $ZZ\to ZZ$ scattering is
not possible in lowest order of the Standard Model, this channel is
relatively clean, though suppressed by the small $eeZ$ initial-state
couplings.  We apply the same cuts on $M_{\rm inv}(ZZ)$, $M_{\rm
recoil}$, and $\cos\theta$ as for the previous channels, and require
both final-state electrons to be detected ($\theta>10^\circ$).  The
resulting cross section is shown as a function of $\alpha_{10}$ in
Fig.\ref{alpha10}.  [$\alpha_{10}$ is actually embedded in the
combination $(\alpha_4+\alpha_5) + 2(\alpha_6+\alpha_7) +
2\alpha_{10}$, yet the parameters $\alpha_4\ldots\alpha_7$ are assumed
to be pre-determined.]  From the $1\sigma$ band of the cross section
we conclude that $|\alpha_{10}|$ can be bounded to less than $\sim
0.002$ at an $e^+e^-$ collider of $1.6\;{\rm TeV}$ for an integrated
luminosity of $500\;{\rm fb}^{-1}$.  The sensitivity is an order of
magnitude better at $1.6\;{\rm TeV}$ than at $800\;{\rm GeV}$.

\section{Conclusions}
\label{sec:conc}
As demonstrated in this analysis, $e^\pm e^-$ linear colliders
operating in the TeV range are able to shed light on the details of
$WW$ scattering even in the most difficult case where no new
resonances are present in the accessible energy range.  The accuracy
of simultaneous measurements of the chiral parameters $\alpha_{4,5}$
will be of the order $0.002$ with an integrated luminosity of
$500\;{\rm fb}^{-1}$.  Furthermore, the $SU(2)_c$-violating quartic
gauge couplings, $\alpha_{6,7,10}$ can be measured directly by
studying all possible $WW$ scattering channels.  Analogous processes
can be studied at the LHC, where a somewhat lower sensitivity on
$\alpha_{4,5}$ is predicted~\cite{Dobado}.  On the other hand, if
there are new resonances in $WW$ scattering below the maximal
accessible energy, they will be observed in different channels at both
the LHC and $e^\pm e^-$ (or $\mu^+\mu^-$)
colliders~\cite{LHC,BCHP,eminus,muon}.

The error with which the reference values 
$\{\alpha_4,\alpha_5\}=\{0,0\}$
of the next-to-leading corrections will be measured, can be
re-interpreted as the error with which the leading amplitudes can be
determined, \emph{i.e.}, the master amplitude $A(s,t,u)_{\rm
LO}=s/F^2$.  At the $e^+e^-$ collider energy $\sqrt{s}=1.6\;{\rm
TeV}$, the scale parameter $F=v$ can be determined to with high accuracy
\begin{equation}
  \Delta F/F \lesssim 5\%
\end{equation}
for an integrated luminosity of $\int{\cal L}=500\;{\rm fb}^{-1}$.
Since the form of this amplitude is characteristic for the chiral
symmetry breaking as the mechanism driving the dynamics of the
strongly interacting $W$~bosons, this test is the most important goal
in analyzing the strong interaction threshold before resonance
phenomena are expected to be observed at still higher energies.  No
dynamical mechanisms other than the Higgs mechanism and spontaneously
broken strong interaction theories have been worked out so far through
which masses of the electroweak gauge bosons could be generated in a
natural way.

\section*{Acknowledgements}
H.J.H\ is grateful to T.~Han, I.~Kuss, and A.~Likhoded for valuable
discussions.  E.B.\ is supported by the Deutsche
Forschungsgemeinschaft (DFG), H.J.H.\ by the Alexander von Humboldt
Stiftung; W.K.\ by the Bundesministerium f\"ur Bildung und Forschung
(BMBF); E.B.\ and A.P.\ acknowledge a grant (96-02-19773a) of the
Russian Foundation of Basic Research (RFBR); C.P.Y.\ a NSF grant
(contract PHY-9507683).

\appendix
\section{Unitarity bounds on \boldmath$\alpha_4,\alpha_5$}
\label{sec:uni}
If custodial $SU(2)_C$ symmetry is assumed, the weak isospin
amplitudes $A^{(I)}$ ($I=0,1,2$) for longitudinal $WW$ scattering in
the asymptotic regime ($|s|,|t|,|u|\gg M_W^2$) are given as follows
\begin{eqnarray}
  A^{(0)} & = & 3A(s ,t ,u )+A(t ,s ,u )+ A(u ,t ,s )\nonumber\\
  A^{(1)} & = &  A(t ,s ,u )-A(u ,t ,s )\nonumber\\
  A^{(2)} & = &  A(t ,s ,u )+A(u ,t ,s)
\end{eqnarray}
The master amplitude $A(s,t,u)$ has been discussed to next-to-leading
order earlier, 
\begin{equation}\label{aZZ-expr-A}
  A(s,t,u) = \frac{s}{v^2} + \alpha_4\frac{4(t^2+u^2)}{v^4}
        + \alpha_5\frac{8s^2}{v^4}
\end{equation}
The isospin amplitudes may be decomposed with respect to orbital
angular momentum according to
\begin{equation}
  A^{(I)} = 32\pi\sum^{\infty}_{\ell =0}
        (2\ell +1)P_\ell (\cos\theta )\,a_{\ell}^I 
\end{equation}
From the parameterization~(\ref{aZZ-expr-A}) the non-zero
amplitudes $a_0^I$ can be extracted:
\begin{eqnarray}
  \mbox{$S$ wave:} \qquad\qquad
  a_0^0 &=& \frac{1}{64\pi}
        \left[+\frac{4s}{v^2} + 
   \frac{16}{3}\left(7\alpha_4+11\alpha_5\right)
                \frac{s^2}{v^4} \right] \\
  a_0^2 &=& \frac{1}{64\pi}
        \left[-\frac{2s}{v^2}+\frac{32}{3}
   \left(2\alpha_4+\alpha_5\right)
                \frac{s^2}{v^4} \right] \\
  \mbox{$P$ wave:} \qquad\qquad
  a_1^1 &=& \frac{1}{64\pi}\left[
        +\frac{2s}{3v^2} + 
         \frac{8}{3}\left(\alpha_4-2\alpha_5\right)
                \frac{s^2}{v^4} \right] \\
  \mbox{$D$ wave:} \qquad\qquad
  a_2^0 &=& \frac{1}{64\pi}\left[
        0 + \frac{16}{15}\left(2\alpha_4+\alpha_5\right)
                \frac{s^2}{v^4} \right] \\
  a_2^2 &=& \frac{1}{64\pi}\left[
        0 + \frac{8}{15}\left(\alpha_4+2\alpha_5\right) 
                \frac{s^2}{v^4} \right] 
\end{eqnarray}
All amplitudes with $I+\ell=\mbox{odd}$ vanish due to CP invariance.
Angular momentum states with $\ell>2$ are populated by higher-order
operators in the chiral expansion.

\begin{figure}
\begin{center}
\hspace*{-3.7cm}
\includegraphics[width=19cm,height=12.9cm]{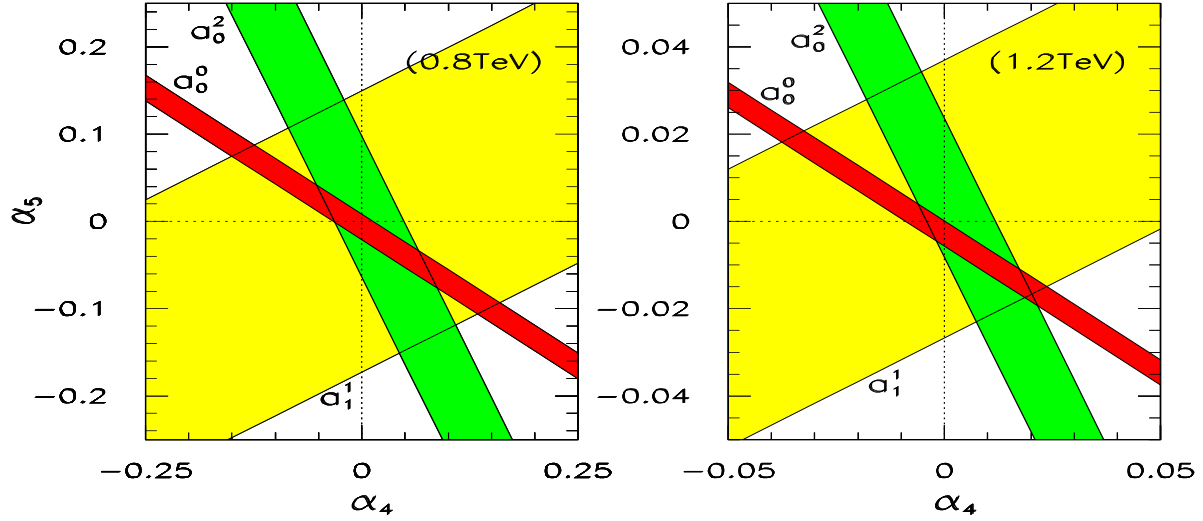}
\end{center}
\vspace*{-6.3cm}
\caption{Region in $\alpha_{4,5}$ allowed by tree-level
unitarity for $WW$ elastic scattering at a subprocess energy of
$0.8\;{\rm TeV}$ (left) resp.\ $1.2\;{\rm TeV}$ (right).}
\label{fig:Uni}
\end{figure}

Two-body elastic unitarity requires $|a_{\ell}^I-\frac{i}{2}|=1/2$.
Once a partial-wave amplitude approaches the limit ${\rm
Re}\,a_{\ell}^I=1/2$, rescattering effects set in which induce a phase
shift that unitarizes the amplitudes.  Such effects can no longer be
described within the effective-theory approach in a model-independent
way.  The validity of the chiral expansion is therefore limited to
$WW$-scattering energies $\sqrt{\hat s}$ and values of the parameters
$\alpha_i$ such that
\begin{equation}
  |a_{\ell}^I|\lesssim 1/2
\end{equation}
In Fig.\ref{fig:Uni} we display the allowed region in the
$[\alpha_4,\alpha_5]$ plane for $\sqrt{\hat s}=0.8\;{\rm TeV}$ and
$1.2\;{\rm TeV}$, which cover the main energy range of the $WW$
scattering subprocess in the analysis.  The strongest limits can be
derived from unitarity in the $S$-wave for isospin~$0$ and~$2$
channels.  The limit from the $I=\ell=1$ channel is significantly
weaker. As demonstrated in Fig.\ref{fig:Uni}, the unitarity bounds are
very sensitive to the energy scale: For $\sqrt{\hat s}= 1.2\;{\rm
TeV}$ they are more stringent by about a factor of $5$ than the bounds
at $0.8\;{\rm TeV}$.  However, they only marginally restrict the
$\alpha_i$ parameters in the range we are interested in
($|\alpha_i|\lesssim 0.005$).  Thus they do not affect the validity of
the chiral expansion in the range considered in the present analysis.

\section{Radiative corrections}
\label{sec:loop}
\begin{figure}
\unitlength1mm
\begin{center}
\newcommand{\rgraph}[2]{%
  \begin{fmfgraph}(30,15)
  \fmfleftn{i}{2} \fmfrightn{o}{2}
  \fmf{phantom}{i1,v1,i2}  \fmf{phantom}{o1,v2,o2}
  \fmf{phantom}{v1,v2}
  \fmffreeze #1 \fmffreeze #2
  \end{fmfgraph}
}
\newcommand{\nrgraph}[2]{%
  \begin{fmfgraph}(20,15)
  \fmfleftn{i}{2} \fmfrightn{o}{2}
  \fmf{phantom}{i1,v1,i2}  \fmf{phantom}{o1,v1,o2}
  \fmffreeze #1 \fmffreeze #2
  \end{fmfgraph}
}
\rgraph{\fmf{plain}{i1,v1,i2}\fmf{plain}{o1,v2,o2}}
        {\fmf{plain,left=.70}{v1,v2,v1}}
\nrgraph{\fmf{plain}{i1,w1,o1}\fmf{plain}{i2,w2,o2}\fmf{phantom}{w1,w2}}
        {\fmf{plain,left=.80}{w1,w2,w1}}
\fmfcmd{%
  style_def over (expr p) = 
    undraw subpath(.2,.8) of p withpen currentpen scaled 6;
    draw p
  enddef;}
\nrgraph{\fmf{plain}{i1,w1}\fmf{plain}{i2,w2}\fmf{phantom}{o1,w1,w2,o2}}
       {\fmf{plain,left=.80}{w1,w2,w1}
         \fmfcmd{draw      vloc __w2{dir 30}..{dir-60}vloc __o1;
                 draw_over(vloc __w1{dir-30}..{dir 60}vloc __o2);}}
\end{center}
\caption{Leading one-loop contributions to the $WW$ scattering
amplitude, expressed in terms of Goldstone-boson scattering.}
\label{1loop-graphs}
\end{figure}
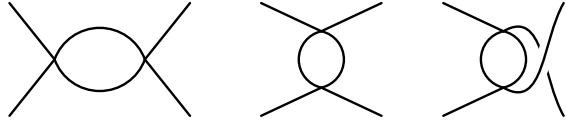

The leading radiative corrections of the tree-level
amplitude~(\ref{aZZ-expr}) are generated by the one-loop corrections
from pure Goldstone dynamics (Fig.\ref{1loop-graphs}).  They give
rise to additional $SU(2)_c$-symmetric contributions of the
form~\cite{1loop}
\begin{equation}\label{aZZ-1loop}
  \Delta A(s,t,u)_{\rm 1\;loop}
  = \frac{1}{16\pi^2v^4}\left\{
      -\frac{(t-u)}{6}\left[t\ln\frac{-t}{\mu^2}-u\ln\frac{-u}{\mu^2}\right]
      -\frac{s^2}{2}\ln\frac{-s}{\mu^2}\right\}
\end{equation}
The real part of these corrections is taken to vanish at the symmetric
point $\mu^2=s=-2t=-2u$, which corresponds to the scattering angle
$\theta= \pi/2$.  Infinities are absorbed in the definition of the
renormalized parameters $\alpha_{4,5}(\mu)$.  A shift in the scale
$\mu$ may be mapped into a finite renormalization of the parameters
$\{\alpha_4,\alpha_5\}$:
\begin{eqnarray}
  \alpha_4(\mu) &=& \alpha_4(\mu_0) 
        - \frac{1}{16\pi^2}\,\frac16\ln\frac{\mu}{\mu_0}\\
  \alpha_5(\mu) &=& \alpha_5(\mu_0) 
        - \frac{1}{16\pi^2}\,\frac1{12}\ln\frac{\mu}{\mu_0}
\end{eqnarray}
The leading-order term $A(s,t,u)_{\rm LO}=s/v^2$ is not renormalized.
The same holds true for the next-to-leading order custodial
$SU(2)_c$-breaking coefficients $\alpha_{6,7,10}$ because standard
one-loop corrections generate only $SU(2)_c$-symmetric amplitudes.

The leading contributions are built up by Goldstone loops since
contributions of transverse $W,Z$ bosons are suppressed by the
electroweak gauge couplings and by reduced enhancement factors in the
energy~\cite{HKY97}.

\begin{figure}
\begin{center}
\includegraphics{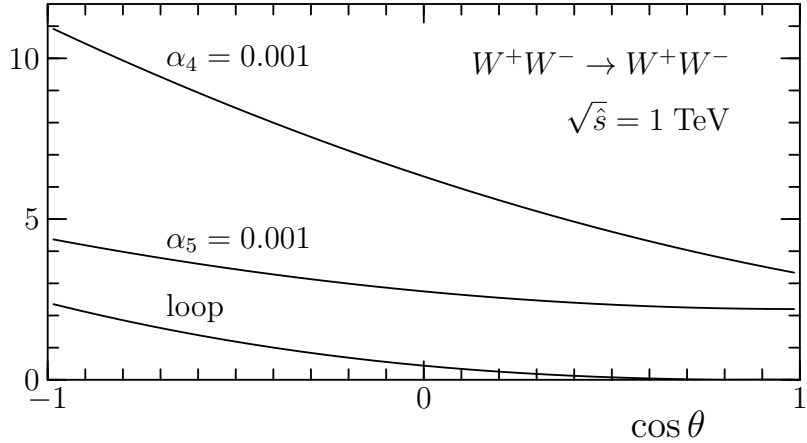}
\end{center}
\vspace*{-5mm}
\caption{Comparison of the leading one-loop corrections $|{\rm
Re}\,(\Delta A)|$ to the longitudinal $W^+W^-\to W^+W^-$ scattering
amplitude, with the effects due to nonvanishing values of $\alpha_4$
resp.\ $\alpha_5$.}
\label{fig:1loop}
\end{figure}

Since the loop corrections~(\ref{aZZ-1loop}) will affect the final
results, it is necessary to estimate their impact.  In
Fig.\ref{fig:1loop} a comparison is presented between the various
contributions to the elastic scattering of longitudinal polarized $W$
bosons, $A(W^+W^-\to W^+W^-)$, as a function of the scattering angle.
The magnitude of the loop corrections, evaluated at the
renormalization point $\mu=\sqrt{\hat s}$, is confronted with the
effects of the next-to-leading order corrections ${\cal L}_4$ and
${\cal L}_5$ on the scattering amplitude.  The loop corrections are
apparently significantly smaller than the chiral contributions for
coefficients $\alpha_4$ and $\alpha_5=0.001$.  Since this is the size
of the sensitivity we are aiming at, cf.\ Fig.\ref{contour1600}, we
can conclude that the longitudinal loop corrections do not invalidate
the previous tree-level results.

\section{Decomposition of helicity amplitudes}
\label{sec:hel}
The partial wave decomposition formula for the helicity amplitudes 
of the process
\begin{equation}
  W^a_{\lambda_1}W^b_{\lambda_2}\to
  W^c_{\lambda_3}W^d_{\lambda_4}
\end{equation}
is defined as~\cite{Hel}
\begin{equation}
  A(\lambda_1\lambda_2,\lambda_3\lambda_4)
  = \exp [i(\lambda -\lambda')\varphi ]\sum_J
        A_J(\lambda_1\lambda_2,\lambda_3\lambda_4)\,
        d^J_{\lambda\lambda'}(\theta )
\end{equation}
where $\lambda \equiv \lambda_1-\lambda_2$, $\lambda'\equiv
\lambda_3-\lambda_4$; and
\begin{eqnarray}
  d^J_{\lambda\lambda'}(\theta )
  &=&  \sum_{s=0}^{J} (-)^s
        \frac{\left[(J+\lambda )!(J-\lambda )!(J+\lambda')!
                (J-\lambda')!\right]^{1/2}}
        {s!(J-s-\lambda )!(J-s+\lambda')!(\lambda -\lambda'+s)! }
	\nonumber\\
  &&	\times
        \left(\cos\frac{\theta}{2}\right)^{2(J-s)+\lambda'-\lambda}
        \left(-\sin\frac{\theta}{2}\right)^{2s+\lambda-\lambda'} 
\end{eqnarray}

Each $2\to 2$ gauge-boson scattering process is described by a total
of $3^4=81$ helicity amplitudes.  However, by applying C,P,T
transformations, they can be reduced to a basic set of 17, 20, and 13
independent amplitudes for the processes $W^+W^-\to W^+W^-$,
$W^+W^-\to ZZ$, and $W^-W^-\to W^-W^-$, respectively, which we present
in tabular form.  In Tab.\ref{tab:hel-a-WW}--\ref{tab:hel-a-ZZ}
the contributions $A_i$ to the individual helicity amplitudes which
are proportional to the NLO coefficients $\alpha_i$ are listed.  In
Tab.\ref{tab:hel-cs-WW}--\ref{tab:hel-t-WW} the $s$-, $t$-channel
exchange, and contact terms are presented in LO for the main
process $W^+W^-\to W^+W^-$.  We use the notation
\begin{equation}
  A_J = 
        \left(\frac{E}{M_W}\right)^{e_L}\left[
        \sum_{V}g_V^2\left(
        \frac{p^2}{s-M_V^2}\hat A_s 
        + \frac{E^2}{t-M_V^2} \hat A_t
        + \frac{E^2}{u-M_V^2} \hat A_u
        \right) + g^2 \hat A_c 
        + g^4\sum_{i=4,5}\alpha_i\hat A_i\right]
\end{equation}
and
\begin{equation}
  \beta\equiv p/E\qquad
  \beta_W\equiv p/E_W \qquad
  \beta_Z\equiv p/E_Z
\end{equation}
where $p=|\vec p\,|$ is the length of 3-momentum of each incoming
$W(Z)$ boson in the c.m.\ frame, and $E_W(E_Z)$ is the corresponding
c.m.\ energy.  When the two incoming gauge bosons have equal masses,
we remove the subscript of $E_W$ or $E_Z$.  The vector boson masses
and couplings are denoted by $M_V,g_V$ ($V=W,Z,\gamma$ referring to
the exchanged particle), where
\begin{equation}
  g_W\equiv g=e/s_w\qquad
  g_Z=e c_w^2/s_w\qquad
  g_\gamma=e
\end{equation}
with $s_w=\sin\theta_w$, $c_w=\cos\theta_w$.  

\begin{table}[T]\small
\begin{displaymath}
\begin{array}{|l||c|c|c||c|c|c|}
\hline
& \multicolumn{3}{c||}{\alpha_4} 
& \multicolumn{3}{c|}{\alpha_5} \\
\hline
& J=0 & 1 & 2 &  0 & 1 & 2 \\ 
\hline\hline
  \hat{A}^J_i(00,00)
& {2(1+\beta^2+2\beta^4)}
& {2\beta^2}
& {2} 
& {\frac{4}{3}(2+3\beta^2+3\beta^4)}
& {-4\beta^2}
& {\frac{4}{3}}  \\ 
\hline
\hat{A}^J_i(+0,00)
& {-}
& {-\beta^2}
& {-\sqrt{3}}
& {-}
& {2\beta^2}
& {-\frac{2}{\sqrt{3}}} \\
\hat{A}^J_i(++,00)
& {-(2+\beta^2)}
& {-}
& {1}
& {-\frac{2}{3}(4+3\beta^2)}
& {-}
& {\frac{2}{3}} \\
\hat{A}^J_i(+0,0+)
& {-}
& {-\frac{1}{2}+2\beta^2}
& {3}
& {-}
& {2}
& {1} \\
 \hat{A}^J_i(+0,+0)
& {-} 
& {-{1\over2}-\beta^2}
& {3\over2}
& {-}
& {1-2\beta^2}
& {1} \\
\hat{A}^J_i(+0,-0)
& {-}
& {-{1\over2}+\beta^2}
& {-{3\over2}}
& {-}
& {1+2\beta^2}
& {-1}  \\
 \hat{A}^J_i(+0,0-)
& {-}
& {-{1\over2}-2\beta^2}
& {-{3\over2}}
& {-}
& {1}
& {-1}  \\
 \hat{A}^J_i(+-,00)
& {-}
& {-}
& {-\sqrt{6}}
& {-}
& {-} 
& {-{2\over3}\sqrt{6}} \\
\hline
 \hat{A}^J_i(++,+0)
& {-}
& {-{1\over2}}
& {\sqrt{3}\over2}
& {-}
& {1}
& {1\over\sqrt{3}}  \\
 \hat{A}^J_i(++,-0)
& {-}
& {1\over2}
& {\sqrt{3}\over2}
& {-}
& {-1}
& {1\over\sqrt{3}}  \\
 \hat{A}^J_i(+-,+0)
& {-}
& {-}
& {3\over\sqrt{2}}
& {-}
& {-}
& {\sqrt{2}} \\
 \hat{A}^J_i(+-,0+)
& {-}
& {-}
& {3\over\sqrt{2}}
& {-}
& {-}
& {\sqrt{2}} \\
\hline
 \hat{A}^J_i(++,++)
& {2}
& {-{1\over2}}
& {1\over2}
& {8\over3}
& {1}
& {1\over3} \\
 \hat{A}^J_i(++,+-)
& {-}
& {-}
& {-{\sqrt{6}\over{2}}}
& {-}
& {-}
& {-{\sqrt{6}\over{3}}} \\
 \hat{A}^J_i(++,--)
& {2}
& {1\over2}
& {1\over2}
& {8\over3}
& {-1}
& {1\over3}  \\
 \hat{A}^J_i(+-,-+)
& {-}
& {-}
& {3}
& {-}
& {-}
& {2} \\
 \hat{A}^J_i(+-,+-)
& {-}
& {-}
& {3}
& {-}
& {-}
& {2}  \\
\hline
\end{array}
\end{displaymath}
\caption{Decomposition of the NLO helicity amplitudes for
$W^+W^-\rightarrow W^+W^-$.}
\label{tab:hel-a-WW}
\end{table}
%
%
\begin{table}[p]\small
\begin{displaymath}
\begin{array}{|c||c|c|c||c|c|c|}
\hline
& \multicolumn{3}{c||}{\alpha_{46}=\alpha_4+\alpha_6} 
& \multicolumn{3}{c|}{\alpha_{57}=\alpha_5+\alpha_7} \\
\hline
& { J=0 }
& { 1 } 
& {2} 
& { 0 }
& { 1 }
& {2} \\ 
\hline\hline
  \hat{A}^J_i(00,00)
& {{2\over3} +2\beta_W^2\beta_Z^2}
& {-}
& {{4\over3}} 
& {2(1+\beta_W^2\beta_Z^2+\beta_W^2+\beta_Z^2)}
& {-}
& {-}  \\ 
\hline
\hat{A}^J_i(+0,00)
& {-}
& {-}
& {-{2\over\sqrt{3}}}
& {-}
& {-}
& {-} \\
\hat{A}^J_i(00,0+)
& {-}
& {-}
& {-{2\over\sqrt{3}}c_w^{-1}}
& {-}
& {-}
& {-} \\
\hat{A}^J_i(++,00)
& {-{2\over3}}
& {-}
& {{2\over3}}
& {-2(1+\beta_Z^2)}
& {-}
& {-} \\
\hat{A}^J_i(00,++)
& {{2\over3}{c_w^{-2}}}
& {-} 
& {{2\over3}c_w^{-2}}
& {-2(1+\beta_Z^2)c_w^{-2}}
& {-}
& {-} \\
\hat{A}^J_i(00,+-)
& {-} 
& {-} 
& {-\frac{4}{\sqrt{6}}c_w^{-2}}
& {-}
& {-}
& {-} \\
\hat{A}^J_i(+-,00)
& {-} 
& {-} 
& {-\frac{4}{\sqrt{6}}} 
& {-}
& {-}
& {-} \\
 \hat{A}^J_i(+0,0-)
& {-} 
& {-c_w^{-1}\beta_W\beta_Z}
& {-c_w^{-1}}
& {-}
& {-}
& {-} \\
\hat{A}^J_i(+0,0+)
& {-}
& {c_w^{-1}\beta_W\beta_Z}
& {c_w^{-1}}
& {-}
& {-}
& {-}  \\
\hline
 \hat{A}^J_i(++,+0)
& {-}
& {-} 
& {-\frac{1}{\sqrt{3}}c_w^{-1}}
& {-}
& {-}
& {-}  \\
 \hat{A}^J_i(+0,++)
& {-}
& {-}
& {-\frac{1}{\sqrt{3}}c_w^{-2}}
& {-}
& {-} 
& {-} \\
 \hat{A}^J_i(++,0-)
& {-}
& {-} 
& {{1\over\sqrt{3}}c_w^{-1}}
& {-}
& {-}
& {-} \\
 \hat{A}^J_i(0-,++)
& {-}
& {-} 
& {{1\over\sqrt{3}} c_w^{-1} }
& {-}
& {-}
& {-} \\
 \hat{A}^J_i(0+,+-)
& {-}
& {-}
& {\sqrt{2}c_w^{-2}}
& {-}
& {-}
& {-} \\
 \hat{A}^J_i(+-,0+)
& {-}
& {-}
& {\sqrt{2}c_w^{-1}}
& {-}
& {-}
& {-} \\
\hline
 \hat{A}^J_i(++,++)
& {{2\over3}c_w^{-2}}
& {-}
& {\frac{1}{3}c_w^{-2}}
& {2 c_w^{-2}}
& {-}
& {-} \\
 \hat{A}^J_i(++,--)
& {{2\over3}c_w^{-2}}
& {-}
& {\frac{1}{3}c_w^{-2}}
& {2 c_w^{-2}}
& {-}
& {-} \\
 \hat{A}^J_i(++,+-)
& {-}
& {-}
& {-{\sqrt{6}\over3}c_w^{-2}}
& {-}
& {-} 
& {-} \\ 
 \hat{A}^J_i(+-,++)
& {-}
& {-}
& {-{\sqrt{6}\over3} c_w^{-2}}
& {-}
& {-}
& {-} \\
 \hat{A}^J_i(+-,-+)
& {-}
& {-}
& { 2 c_w^{-2} }
& {-}
& {-}
& {-}  \\
\hline
\end{array}
\end{displaymath}
\caption{Decomposition of the NLO helicity amplitudes for $W^+W^-\to ZZ$.}
\label{tab:hel-a-ZZ}
\end{table}

\begin{table}[p]\small
\begin{displaymath}
\begin{array}{|c||c|c|c||c|c|c|}
\hline
& \multicolumn{3}{c||}{\alpha_4} 
& \multicolumn{3}{c|}{\alpha_5} \\
\hline
& { J=0 }
& { 1 } 
& {2} 
& { 0 }
& { 1 }
& {2} \\ 
\hline\hline
  \hat{A}^J_i(00,00)
& {4(\frac{2}{3}+\beta^2+\beta^4)}
& {-}
& {{4\over3}} 
& {4(\frac{1}{3}+\beta^4)}
& {-}
& {\frac{8}{3}}  \\ 
\hline
\hat{A}^J_i(+0,00)
& {-}
& {-}
& {-\frac{2}{\sqrt{3}}}
& {-}
& {-}
& {\frac{4}{\sqrt{3}}} \\
\hline
\hat{A}^J_i(++,00)
& {-\frac{2}{3}(4+3\beta^2)}
& {-}
& {-\frac{2}{3}}
& {-\frac{4}{3}} 
& {-}
& {\frac{4}{3}} \\
\hat{A}^J_i(+0,0+)
& {-}
& {\beta^2}
& {1}
& {-}
& {2\beta^2}
& {2} \\
\hat{A}^J_i(00,+-)
& {-}
& {-} 
& {-\frac{4}{\sqrt{6}}}
& {-} 
& {-} 
& {-\frac{8}{\sqrt{6}}} \\
 \hat{A}^J_i(+0,0-)
& {-}
& {-\beta^2} 
& {-1}
& {-}
& {-2\beta^2}
& {-2}  \\
\hline
 \hat{A}^J_i(++,+0)
& {-}
& {-}
& {-\frac{1}{\sqrt{3}}} 
& {-}
& {-}
& {-2\over\sqrt{3}}  \\
 \hat{A}^J_i(++,0-)
& {-}
& {-} 
& {1\over\sqrt{3}}
& {-}
& {-}
& {2\over\sqrt{3}}  \\
 \hat{A}^J_i(0+,+-)
& {-}
& {-}
& {\sqrt{2}}
& {-}
& {-}
& {2\sqrt{2}} \\
\hline
 \hat{A}^J_i(++,++)
& {\frac{8}{3}}
& {-}
& {1\over3}
& {3\over2}
& {-}
& {2\over3} \\
 \hat{A}^J_i(++,+-)
& {-}
& {-}
& {-{\sqrt{6}\over{3}}}
& {-}
& {-}
& {-{2\sqrt{6}\over{3}}} \\
 \hat{A}^J_i(++,--)
& {8\over3}
& {-}
& {1\over3}
& {4\over3}
& {-}
& {2\over3}  \\
 \hat{A}^J_i(+-,-+)
& {-}
& {-}
& {2}
& {-}
& {-}
& {4} \\
\hline
\end{array}
\end{displaymath}
\caption{Decomposition of the NLO helicity amplitudes for $W^\pm W^\pm\to
W^\pm W^\pm$.} 
\label{tab:hel-a-sWW}
\end{table}
\begin{table}[p]\small
\begin{displaymath}
\begin{array}{|c||c|c|c||c|c|c|}
\hline 
&  \multicolumn{3}{|c||}{\mbox{Contact graph}} 
&  \multicolumn{3}{|c|}{\mbox{$s$-channel $Z/\gamma$--exchange}}\\
\hline
& { J=0 }
& { 1 } 
& {2} 
& { 0 }
& { 1 }
& {2} \\ 
\hline\hline
  \hat{A}^J_i(00,00)
& {-\frac{2}{3}(1+\beta^2)} 
& {6\beta^2}
& {\frac{2}{3}} 
& {-} 
& {-4(3-\beta^2)^2}
& {-}  \\ 
\hline
\hat{A}^J_i(+0,00)
& {-}
& {-3\beta^2}
& {-\frac{1}{\sqrt{3}} }
& {-}
& {8(3-\beta^2)}
& {-} \\
\hat{A}^J_i(++,00)
& {\frac{2}{3}+\beta^2}
& {-}
& {1\over3}
& {-} 
& {4(\beta^2-3)}
& {-} \\
\hat{A}^J_i(+0,0+)
& {-}
& {-\frac{3}{2}+2\beta^2}
& {1\over2}
& {-}
& {-16}
& {-} \\
 \hat{A}^J_i(+0,+0)
& {-} 
& {-{3\over2}+\beta^2}
& {1\over2}
& {-}
& {-16}
& {-} \\
\hat{A}^J_i(+0,-0)
& {-}
& {-{3\over2}-\beta^2}
& {-{1\over2}}
& {-}
& {16}
& {-}  \\
 \hat{A}^J_i(+0,0-)
& {-}
& {-{3\over2}-2\beta^2}
& {-{1\over2}}
& {-}
& {16}
& {-}  \\
 \hat{A}^J_i(+-,00)
& {-}
& {-}
& {\frac{\sqrt{6}}{3}}
& {-}
& {-}  
& {-} \\ 
\hline
 \hat{A}^J_i(++,+0)
& {-}
& {{3\over2}}
& {-\frac{\sqrt{3}}{6}}
& {-}
& {8}
& {-} \\ 
 \hat{A}^J_i(++,-0)
& {-}
& {3\over2}
& {\sqrt{3}\over6}
& {-}
& {-8}
& {-} \\  
 \hat{A}^J_i(+-,+0)
& {-}
& {-}
& {1\over\sqrt{2}}
& {-}
& {-}
& {-} \\
 \hat{A}^J_i(+-,0+)
& {-}
& {-}
& {1\over\sqrt{2}}
& {-}
& {-}
& {-} \\
\hline
 \hat{A}^J_i(++,++)
& {-\frac{2}{3}}
& {-{3\over2}}
& {1\over6}
& {-}
& {-4}
& {-} \\
\hline
 \hat{A}^J_i(++,+-)
& {-}
& {-}
& {-\frac{1}{\sqrt{6}}}
& {-}
& {-} 
& {-} \\  
 \hat{A}^J_i(++,--)
& {-\frac{2}{3}}
& {3\over2}
& {1\over6}
& {-}
& {-4}
& {-}  \\
 \hat{A}^J_i(+-,-+)
& {-}
& {-}
& {1}
& {-}
& {-}
& {-} \\
 \hat{A}^J_i(+-,+-)
& {-}
& {-}
& {1}
& {-}
& {-}
& {-}  \\
\hline
\end{array}
\end{displaymath}
\caption{Decomposition of the NLO helicity amplitudes for $W^+W^-\to W^+W^-$.}
\label{tab:hel-cs-WW}
\end{table}

\begin{table}[p]\small
\begin{displaymath}
\begin{array}{|c||c|c|c|c|}
\hline 
& \multicolumn{4}{|c|}{
\mbox{$t$-channel $Z/\gamma$-exchange}} \\
\hline
& { J=0 }
& { 1 } 
& {2} 
& { 3 } \\
\hline\hline
  \hat{A}^J_t(00,00)
& { -10(2+\beta^2+4\beta^4+\beta^6)} 
& { 6\beta^2(-23+50\beta^2-5\beta^4)}
& { 20(-2+11\beta^2-10\beta^4)} 
& { -12\beta^2} \\
\hline
\hat{A}^J_t(+0,00)
& {-}
& { 18\beta^2(7-5\beta^2)}
& { 10\sqrt{3}(2-9\beta^2+5\beta^4) }
& { 4\sqrt{6}\beta^2}\\
\hat{A}^J_t(++,00)
& { 20(2-3\beta^2+2\beta^4)}
& { -114\beta^2}
& { 20(-2+9\beta^2-2\beta^4)}
& { -6\beta^2} \\
\hat{A}^J_t(+0,0+)
& {-}
& { -6(5+17\beta^2+5\beta^4)}
& { 10(-3+13\beta^2-3\beta^4)}
& { -8\beta^2} \\
 \hat{A}^J_t(+0,+0)
& {-} 
& { 3(-10-39\beta^2+25\beta^4)}
& { 5(-6+25\beta^2-3\beta^4)}
& { -8\beta^2} \\
\hat{A}^J_t(+0,-0)
& {-}
& { 3(-10+39\beta^2-5\beta^4)}
& { 5(6-17\beta^2+3\beta^4)}
& { 8\beta^2}  \\
 \hat{A}^J_t(+0,0-)
& {-}
& { 6(-5+12\beta^2-5\beta^4)}
& { 10(3-8\beta^2+3\beta^4)}
& { 8\beta^2}  \\
 \hat{A}^J_t(+-,00)
& {-}
& {-}
& { 10\sqrt{6}(2-5\beta^2+2\beta^4) }
& { 2\sqrt{30}\beta^2} \\ 
\hline
 \hat{A}^J_t(++,+0)
& {-}
& { 6(5+13\beta^2)}
& { 10\sqrt{3}(1-5\beta^2)}
& { 2\sqrt{6}\beta^2} \\ 
 \hat{A}^J_t(++,-0)
& {-} 
& { 6(5-8\beta^2)}
& { 10\sqrt{3}(-1+2\beta^2)}
& { -2\sqrt{6}\beta^2} \\  
 \hat{A}^J_t(+-,+0)
& {-}
& {-}
& { 10\sqrt{2}(-3+4\beta^2)}
& { -4\sqrt{5}\beta^2} \\
 \hat{A}^J_t(+-,0+)
& {-}
& {-}
& { 10\sqrt{2}(-3+5\beta^2)}
& { -4\sqrt{5}\beta^2} \\
\hline
 \hat{A}^J_t(++,++)
& { -5(4+19\beta^2)}
& { -3(10+9\beta^2)}
& { 5(-2+13\beta^2)}
& { -3\beta^2} \\
 \hat{A}^J_t(++,+-)
& {-}
& {-}
& { 5\sqrt{6}(2-3\beta^2)}
& { \sqrt{30}\beta^2} \\  
 \hat{A}^J_t(++,--)
& { -5(4+\beta^2)}
& { 3(10+\beta^2)}
& { 5(-2+\beta^2)}
& { -3\beta}  \\
 \hat{A}^J_t(+-,-+)
& {-}
& {-}
& { -10(6+\beta^2)}
& { -10\beta^2} \\
 \hat{A}^J_t(+-,+-)
& {-}
& {-}
& { -10(6+5\beta^2)}
& { -10\beta^2}  \\
\hline
\end{array}
\end{displaymath}
\caption{Decomposition of the leading order helicity amplitudes 
for $W^+W^-\to W^+W^-$.}
\label{tab:hel-t-WW}
\end{table}
\clearpage


\end{fmffile}
\end{document}